\documentclass[aps,amsmath,amssymb,showpacs,pre,superscriptaddress]{revtex4-1}
\usepackage{graphicx,color}
\usepackage{amsmath}
\usepackage{enumerate}
\usepackage{mathbbol}
\usepackage{amsfonts}
\usepackage{natbib}

\usepackage{bm}
\usepackage[T1]{fontenc}
\usepackage[latin9]{inputenc}
\usepackage{geometry}
\usepackage{amsmath,amssymb}
\usepackage{graphicx}
\usepackage{esint}

\usepackage{color}

\begin{document}

\title{Exact extreme value statistics at mixed order transitions}

\author{Amir Bar}
\affiliation{Department of Complex Systems, Weizmann Institute, Rehovot, Israel}
\author{Satya N. \surname{Majumdar}}
\affiliation{Univ. Paris-Sud, CNRS, LPTMS, UMR 8626, Orsay F-91405, France}
\author{Gr\'egory \surname{Schehr}}
\affiliation{Univ. Paris-Sud, CNRS, LPTMS, UMR 8626, Orsay F-91405, France}
\author{David Mukamel}
\affiliation{Department of Complex Systems, Weizmann Institute, Rehovot, Israel}

\begin{abstract}
We study extreme value statistics (EVS) 
for spatially extended models exhibiting mixed order phase transitions
(MOT). These are phase transitions which exhibit features common to
both first order (discontinuity of the order parameter) and second
order (diverging correlation length) transitions. We consider here the truncated inverse distance squared Ising (TIDSI)
model which is a prototypical model exhibiting MOT, and study analytically 
the extreme value statistics of the domain lengths. The lengths of the domains are identically distributed random variables
except for the global constraint that their sum equals the total system size $L$.
In addition, the number of such domains is also a fluctuating variable,
and not fixed. In the paramagnetic phase, we show that the distribution of the largest
domain length $l_{\max}$ converges, in the large $L$ limit, to a Gumbel distribution. 
However, at the critical point (for a certain range of parameters) and in the ferromagnetic
phase, we show that the fluctuations of $l_{\max}$ are governed by novel distributions which 
we compute exactly. Our main analytical results are
verified by numerical simulations.  
\end{abstract}


\maketitle

\tableofcontents

\section{Introduction}

Extreme events are generally rare, but their implications may be of
major importance. Hence, the theory of such events has found many applications
in diverse fields such as geology ({\it e.g.}, earth-quakes analysis), economy
({\it e.g.},~stock market fluctuations), physics ({\it e.g.}, properties of ground states of disordered systems) or
biology ({\it e.g.}, evolution theory). The theory of extreme value statistics (EVS) for independent and identically distributed (i.i.d.) random
variables is well known since the work of
Tippett, Fisher, Fr\'echet, Gumbel, Weibull and others~\cite{fisher1928limiting,frechet1927loi,gnedenko1943distribution,gumbel1958statistics,weibull1951wide}.
However, the study of extreme values for sets of correlated variables
is an active field of research (for a recent review see \cite{majumdar2014extreme}).
In this work we study a specific class of correlated variables,
which represent degrees of freedom of a spatially extended system poised in a rather
unconventional type of critical point, named mixed order phase transition (MOT).

MOTs are phase transitions in which the order parameter
changes discontinuously, as in first order transitions, but exhibit
diverging correlation length and scale free distributions as in continuous
transitions. Such transitions appear in several distinct contexts including 
one-dimensional Ising model with long range interactions \cite{thouless1969long,yuval1970exact,aizenman1988discontinuity},
models of DNA denaturation \cite{PS1966,KMP2000}, wetting and
depinning transitions \cite{fisher1984walks,blossey1995diverging}, models for
glass and jamming transitions \cite{gross1985mean,toninelli2006jamming,schwarz2006onset,liu2012core}, complex network evolution \cite{liu2012extraordinary,zia2012extraordinary,tian2012nature,bizhani2012discontinuous} or active biopolymer gels \cite{alvarado2013molecular,sheinman2015anomalous}.
While it is clear that these transitions do not fall into the ordinary
classification scheme of phase transitions, there is currently no
theoretical framework which provides a comprehensive classification
of such transitions. One clear distinction between  different MOTs
is the behavior of the correlation length near the transition: in
some cases its divergence is polynomial in the control parameter
({\it e.g.} in the Poland-Scheraga model \cite{PS1966} and in the No-Enclave Percolation model
\cite{sheinman2015anomalous}), while in others the correlation length
exhibits an essential singularity, in the form of stretched exponential
divergence ({\it e.g.} in the Inverse Distance Squared Ising model \cite{thouless1969long} or in 
the Spiral model for jamming \cite{toninelli2006jamming}). Recently
\cite{bar2014mixed,bar2014mixed2}, this distinction in the behavior
of the correlation length was studied in a one-dimensional setting,
using renormalization group (RG) analysis to study, on the same footing,
models from both classes. It is an ongoing research task to find other
relations and distinctions between such transitions, or at least a
framework in which they can be analyzed together. Here we highlight
EVS as unifying concepts for such transitions.

In the examples for MOT mentioned above, mixed order transitions separate
a phase composed of microscopic domains from a phase in which
a macroscopic domain exists. For instance, in the context of DNA denaturation,
the relevant domains are denatured regions, and the MOT involves the
appearance of a macroscopic denatured region. In the context of network
evolution, the relevant domains are the connected components, and
a macroscopic domain is a spanning cluster. The transition itself
can be identified with the change in the scaling of the maximal domain
size with the total system size: the largest domain is sub-extensive
below the transition, and extensive above it. Hence, it is natural
in this context to study the extreme value statistics (EVS) of the
set of domain sizes. 

Obtaining exact results for EVS of generally correlated variables,
such as domain sizes in a network evolution model, is notoriously
hard. In order to gain analytical insight into this problem we study
it in the context of the truncated inverse distance squared Ising model (TIDSI), 
which was introduced in \cite{bar2014mixed} as a bridge
between models exhibiting MOT in one dimension. The sizes of domains
in a configuration of the TIDSI model are essentially independent
variables, apart for a sum constraint which generates correlations [see Eq. (\ref{eq:jpdf}) below].
In addition, the number of domains is fluctuating. Due to this special
structure, many properties of this model, including the extreme value theory
of its domain sizes, are analytically accessible. We find that the
EVS distribution can be either standard independent-variables distribution
or novel EVS distributions, depending on control parameters of the model. 



In this paper we derive the EVS of the TIDSI model analytically and discuss
its important features. The paper is organized as follows. In section II we introduce the TIDSI model, discuss its various 
representations and recall its phase diagram.
In section III we discuss the extreme value theory of the TIDSI, which
is the main result of this paper. In section IV we discuss the direct
relation between the TIDSI and other one-dimensional models which exhibit MOT. Finally we discuss our findings in section V. For completeness,
we review basic results for EVS of i.i.d. random variables in Appendix A. Some technical details have been relegated in Appendix B and C.

\section{The model}

\begin{figure}
\centering
\includegraphics[width=0.7\linewidth]{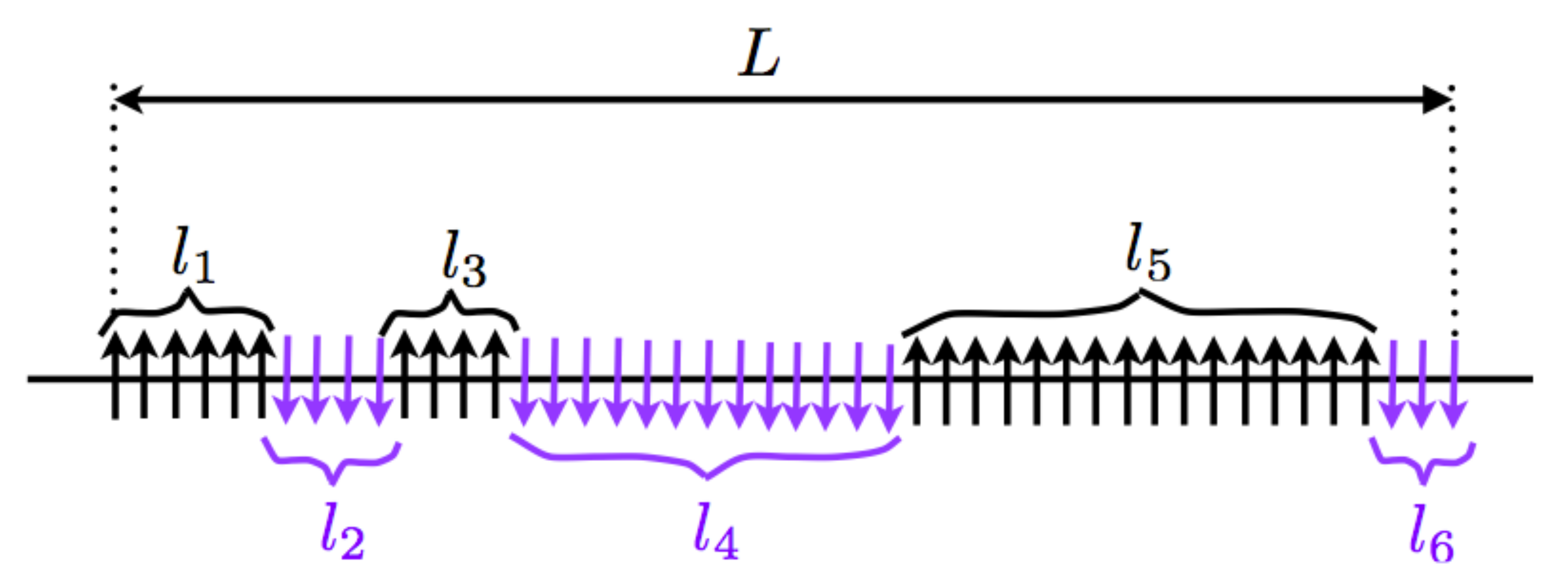}
\caption{Domain representation of the TIDSI model of size $L$. In this configuration, the number of domains is $N=6$. We recall that the interaction is restricted to spins within the same domain. The Boltzmann-Gibbs weight of such a configuration is given by Eq. (\ref{eq:jpdf}). In this paper we study the statistics of the largest domain length $l_{\max} = \max_{1\leq i \leq N} l_i$.}\label{fig:spins}
\end{figure}

The TIDSI model was introduced
in \cite{bar2014mixed} and further analyzed in \cite{bar2014mixed2}.
Originally, the TIDSI was defined as an Ising spin chain with specific
long range interactions. However, in this paper we will focus on its
representation in terms of spin domains (see Fig. \ref{fig:spins}), in the regime in which the
relevant domains are large and hence terms inversely proportional
to domain length can be neglected. We start by reminding the readers the original
TIDSI model in the spin representation and then derive its domain representation. 

\vspace*{0.5cm}

{\bf Spin representation.} In its spin chain representation the TIDSI model is defined on a spin
chain of size $L$. At each site there is an Ising spin $\sigma_{i}=\pm1$. 
There is a standard nearest neighbor ferromagnetic interaction between spins. In addition,
there is a ferromagnetic long range interaction between spins belonging to the same domain, where a domain is a consecutive
set of spins of the same sign (see Fig. \ref{fig:spins}). Thus the long range interaction is truncated by the finite domain size. 
We consider the case where the long range interaction decays asymptotically
according to an inverse quadratic law. The full Hamiltonian of the system in the spin representation thus reads
\begin{eqnarray}
\mathcal{H} & = & -J_{NN}\sum_{i=1}^{N-1}\sigma_{i}\sigma_{i+1}-\sum_{i<j}J(i-j)\sigma_{i}\sigma_{j}\prod_{k=i}^{j-1}\frac{1+\sigma_{k}\sigma_{k+1}}{2},\label{eq_TIDSI:SC_Hamiltonian1}\\
J(r) & \approx & Cr^{-2} \;, \; r \gg 1 \;.\label{eq_TIDSI:SC_Hamiltonian2}
\end{eqnarray}
The product in the second term of Eq. (\ref{eq_TIDSI:SC_Hamiltonian1}) ensures that the long range interaction is restricted
to spins within the same domain. We consider here free boundary conditions. 

\vspace*{0.5cm}

{\bf Domain representation.} A typical configuration of the system will consist of alternating spin domains characterized by sizes $\{l_1, l_2, \cdots, l_N\}$ (see Fig.~\ref{fig:spins}) where
$N$ is the number of domains, which may also vary from configuration to configuration. Note that the variables $l_i$'s satisfy the constraint 
\begin{eqnarray}
\sum_{i=1}^N l_i = L \label{eq_TIDSI:sum_rule}
\end{eqnarray}
where $L$ is the system size. In terms of these domains, the Hamiltonian can be
re-expressed as 
\begin{eqnarray}
\mathcal{H} & = & \sum_{n=1}^{N}\mathcal{H}_{n} - J_{NN} \label{eq_TIDSI:SC_Htot} \;,\\
\mathcal{H}_{n} & = & -J_{NN}\left(l_{n}-2\right)-\sum_{r=1}^{l_{n}}\left(l_{n}-r\right)J(r) \;. \label{eq_TIDSI:SC_Hdom}
\end{eqnarray}
For any long range interaction $J(r)$ that satisfies $r^{2}J\left(r\right)\rightarrow C$
as $r\rightarrow\infty$, we have 
\begin{eqnarray*}
\sum_{r=1}^{l_{n}}J(r) & = & a-\frac{C}{l_{n}}+{\cal O}\left(l_{n}^{-2}\right)\\
\sum_{r=1}^{l_{n}}rJ(r) & = & b+C\log l_{n}+{\cal O}\left(l_{n}^{-1}\right) \;.
\end{eqnarray*}
 For large enough domains we can ignore ${\cal O}\left(l_{n}^{-1}\right)$
corrections. As we will see later, this is justified near the critical point where the domains are typically very
large. Using the sum rule $\sum_{i=1}^N l_i = L$, the linear term $-\left(J_{NN}+a\right)l_{n}$, summed over $n$,
just becomes a constant and hence can be dropped. Hence, under the
approximation of long domains the Hamiltonian is re-expressed as 
\begin{equation}
\mathcal{H}=C\sum_{n}\log l_{n}+\Delta N.\label{eq_TIDSI:Hamiltonian}
\end{equation}
Here $C$ is a constant parameter and $\Delta=2J_{NN}+C+b$ serves as a chemical
potential for the number of domains.

In this domain representation, a configuration ${\cal C}$ of the system is specified by the domain sizes $\{l_1, l_2, \cdots, l_N\}$ and
the number $N$ of domains (see Fig. \ref{fig:spins}). The Boltzmann weight associated with such a configuration ${\cal C}$ is simply $P({\cal C}) \propto e^{-\beta {\cal H}}$, where $\beta = 1/(k_B\,T)$ is the inverse temperature and the Hamiltonian ${\cal H}$ is given in Eq. (\ref{eq_TIDSI:Hamiltonian}). This then leads to the following joint distribution of the domain lengths and their number
\begin{eqnarray}
P(l_1, l_2, \cdots, l_N,N|L) = \frac{1}{Z(L)} \prod_{n=1}^N \frac{e^{-\beta \Delta}}{l_n^c} \, {\delta}_{\sum_{n=1}^N l_n, L} \,, \label{eq:jpdf}
\end{eqnarray}
where $\delta_{i,j}$ is the usual Kronecker delta and $c = \beta C$. For the purpose of the normalization of the full joint distribution, we need here $c > 1$. The normalization constant $Z(L)$ is the partition function given by
\begin{eqnarray}
Z(L) = \sum_{N=1}^\infty \sum_{l_1=1}^\infty \cdots  \sum_{l_N=1}^\infty \prod_{n=1}^N \frac{e^{-\beta \Delta}}{l_n^c} \, {\delta}_{\sum_{n=1}^N l_n, L} \;. \label{eq:partition_func}
\end{eqnarray}
The two natural parameters in the model are the inverse temperature $\beta$ and the fugacity $e^{-\beta \Delta}$. For convenience, we will use an alternative parameterization in terms of the exponent $c$ characterizing the power law decay of the domain size distribution and the temperature $T$. We note that this model has close similarity to the Poland-Scheraga model of DNA denaturation where the number of loops $N$ is also a variable (see later in section \ref{section:other} for discussions). 

\vspace*{0.5cm}

\begin{figure}[hh]
\centering{}\includegraphics[scale=0.5]{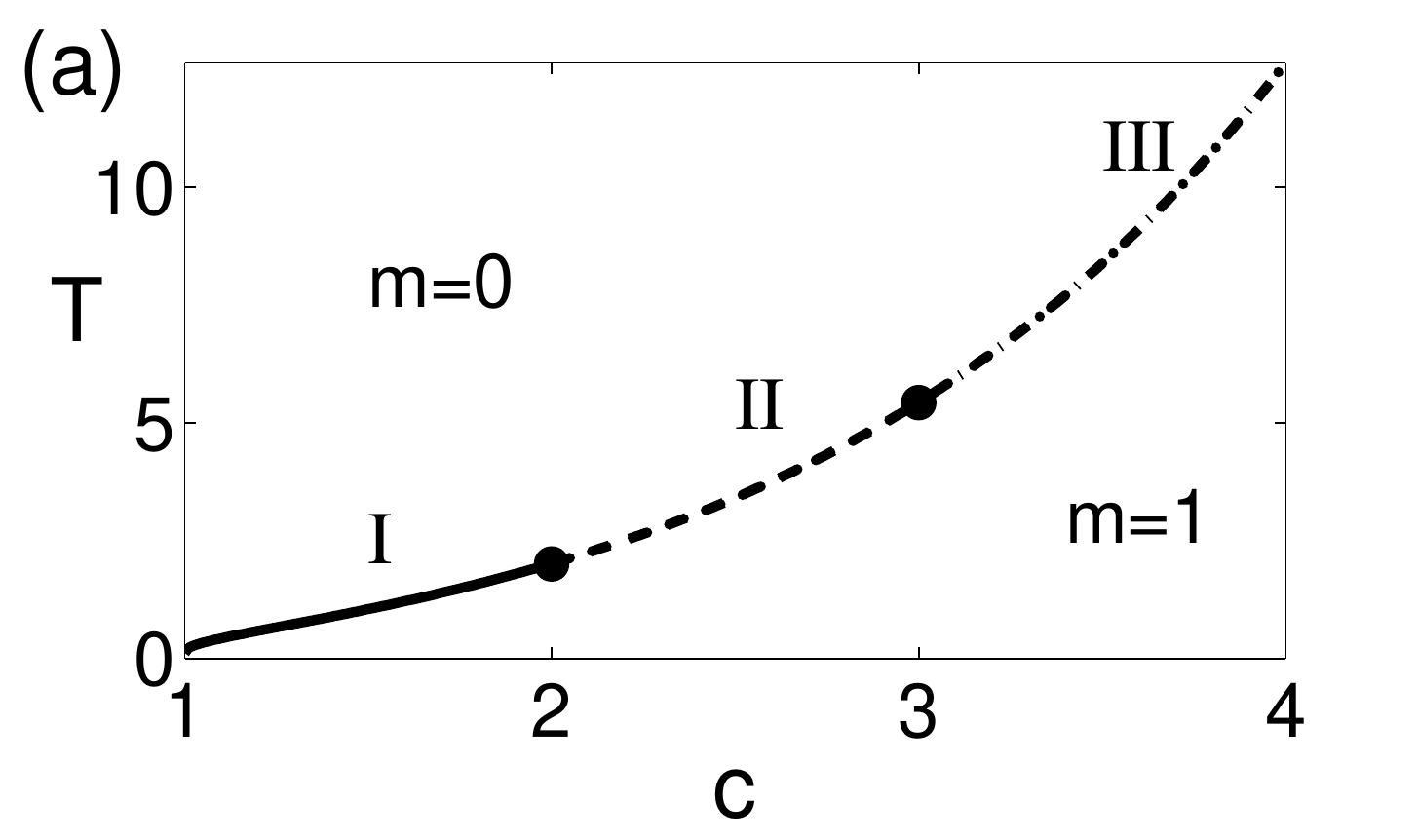}\protect\caption{\label{fig:Phase_diagram}Phase diagram of the model (\ref{eq_TIDSI:Hamiltonian})
in the $\left(c,T\right)$ plane, with $\Delta=1$. The different
regions of the critical line (I-III) are explained in the text.}
\end{figure}

{\bf Phase diagram.}
The phase diagram of the TIDSI model in the $(c,T)$ plane was derived in \cite{bar2014mixed2} (see Fig. \ref{fig:Phase_diagram}).
For completeness we briefly summarize the main results (for zero magnetic
field). There are two relevant order parameters for the TIDSI model:
the density of domains $\rho=\frac{N}{L}$, and the magnetization $m=\sum_{n\geq 1}\left(-1\right)^{n}l_{n}$ \cite{bar2014mixed2}.
For any $C$ and $\Delta>0$, a phase transition is predicted at some
$T_{c}\left(C,\Delta\right)$ which is given by
\begin{equation}
\zeta\left(\beta_{c}C\right)=e^{\beta_{c}\Delta} \;.\label{eq_TIDSI:Tc}
\end{equation}
Here $\beta_{c}=k_{B}T_{c}^{-1}$ and $\zeta\left(\gamma\right) = \sum_{n=1}^\infty n^{-\gamma}$
is the Riemann zeta function. In Fig.~\ref{fig:Phase_diagram} the phase diagram is presented in the $\left(c,T\right)$ plane. 
The critical line in Fig.~\ref{fig:Phase_diagram} separates a high
temperature paramagnetic phase, in which $m=0$ and $\rho>0$, and a
low temperature ferromagnetic phase in which $m=\pm1$ and $\rho=0$.
The ferromagnetic phase is a \emph{condensed }phase, in which all
but a sub-extensive part of the system is in a single macroscopic domain.
The critical line has three different regimes: In regime I ($1<c\le2$)
the magnetization jumps from $0$ to $\pm1$ abruptly, while the density
of domains drops continuously to $0$ in the paramagnetic phase as one approaches the critical line. In
regimes II ($2<c\le3$) and III ($c>3$) both $m$ and $\rho$ change
discontinuously. In regime II the magnetic susceptibility diverges
at the transition, while in regime III it is finite (see \cite{bar2014mixed2}
for definition of magnetic susceptibility in this context). In all
regimes the spin-spin correlation length diverges, and hence the transition
is critical. 

In this paper, the main focus is not on the thermodynamics of this model but rather
on the statistics of the largest domain $l_{\max}$. As discussed above, this is a natural observable 
since the transition from paramagnetism to ferromagnetism occurs via the emergence of
a macroscopic domain as one crosses the critical line. We show in this paper that the statistics
of the largest domain indeed has an extremely rich and novel behavior in different regions of the phase
diagram in the $(c,T)$ plane.

\section{Statistics of the largest domain}

In the domain representation the system is characterized by a fluctuating number of domains $N$ with
domain lengths $l_i$'s distributed according to the joint probability density function (PDF) given in Eq. (\ref{eq:jpdf}). We define
the largest domain size as $l_{\max} = \max_{1 \leq n \leq N} l_n$. This is clearly a random variable and we are interested in computing
its PDF, in particular in the thermodynamic limit $L \to \infty$, in the various regions of the phase diagram in the $(c,T)$ plane. Note that due to (i) the presence of the global constraint $\sum_{n=1}^N l_n = L$ in Eq. (\ref{eq:jpdf}) and (ii) the fluctuating number of domains $N$, the variables $l_n$'s are correlated and therefore the standard extreme value statistics (EVS) of uncorrelated variables are not valid here. Indeed we will see that these two facts lead to results for $l_{\max}$ that are rather different from and richer than the standard EVS results.     

At this point, it is useful to point out that $l_{\max}$ has been recently studied in models that are similar but not exactly identical
to the present case. For example, in the case of zero range process (ZRP), the joint distribution of the number of particles at different sites
of a lattice of size $N$ has a similar structure as in Eq. (\ref{eq:jpdf}):
\begin{eqnarray}\label{eq:ZRP}
P_{\rm ZRP}(l_1,\cdots,l_N|L) \propto \prod_{n=1}^N \frac{1}{l_n^c} \, \delta_{\sum_{n=1}^N l_n,L} \;,
\end{eqnarray} 
where $l_n$ represents the number of particles at site $n$ and $L$ represents the total number of particles. The statistics of $l_{\max}$ in this case (\ref{eq:ZRP}) has been studied in Ref. \cite{evans2008condensation}. Even though structurally Eq. (\ref{eq:ZRP}) is similar to the joint PDF in Eq.~(\ref{eq:jpdf}), there are two important differences: (i) the number of sites $N$ is fixed and hence (ii) there is no explicit fugacity. 

Similarly, the largest time interval between
returns to the origin has been studied for one-dimensional lattice random walks \cite{FIK95} and more generally for renewal processes \cite{GMS2009,GMS2015}. In this case, the joint PDF of the intervals between renewals is given by
\begin{eqnarray}\label{eq:renewal}
P_{\rm REN}(l_1,\cdots,l_N,N|L)  \propto \left[\prod_{n=1}^{N-1} f(l_n)\right] q(l_N) \, \delta_{\sum_{n=1}^N l_n,L} \;,
\end{eqnarray}
where $l_n$'s represent the intervals between renewal events and $L$ represents the total time interval. Here the number of intervals $N$ is a variable as in Eq. (\ref{eq:jpdf}). However, unlike in Eq. (\ref{eq:jpdf}), the first $N-1$ intervals have the same weight $f(l_n)$ but the last one has a different weight $q(l_N) = \sum_{l=l_N+1}^\infty f(l)$. In addition, there is no explicit fugacity as in Eq. (\ref{eq:jpdf}). In these renewal processes (\ref{eq:renewal}), the weight $f(l)$ is taken as an input in the model. In contrast, in the TIDSI model the renewal structure along with the weight $f(l) \propto l^{-c}$ emerge naturally from the Boltzmann weight of an underlying Hamiltonian. Hence the joint PDF in Eq. (\ref{eq:jpdf}) has a richer structure 
as it can be studied in the various regions of the parameter space in the $(c,T)$ plane. Consequently, we will see that the results for the statistics of $l_{\max}$ in the TIDSI model also have a richer structure as summarized below.

\subsection{Summary of main results}

Our main results concern the exact expression for the cumulative distribution of the largest domain,
$P_{1}\left(x\right)=\Pr.\left(l_{\max} \leq x\right)$, or equivalently for its distribution $p_1(x|L) = \Pr.(l_{\max} = x)$,  
in the large $L$ limit, in the various regions of the phase diagram in the ($c,T$) plane (see Fig. \ref{fig:Phase_diagram} and Table \ref{tab1}).

\begin{itemize}
\item $T>T_{c}$: in the paramagnetic phase, the marginal distribution of the domain size has an exponential tail $P(l) \sim l^{-c} e^{-l/\xi}$ where $\xi \equiv \xi(T,c)$ is the typical domain size, which is finite for $T<T_c$ \cite{bar2014mixed,bar2014mixed2}. In this case, we show that the maximal domain size $l_{\max}$, properly shifted and scaled, is distributed according to a Gumbel distribution (see also Fig. \ref{Fig:plot_para} below):
\begin{eqnarray}
P_{1}\left(x|L\right) & \approx & \exp\left[-\exp\left(-\left(\frac{x-b_{L}}{a}\right)\right)\right],\label{eq_TIDSI:P1_TgtTc}
\end{eqnarray}
where $b_L$, which depends explicitly on $L$, and $a$, which is independent of $L$, are given by
\begin{eqnarray}\label{expr:a}
a =-\frac{1}{\log\left({z}^{*}\right)} \;, \; b_{L} = a\,\log\left[\frac{z^* \, L}{(1-z^*){\rm Li}_{c-1}\left(z^{*}\right)\left(a\log L\right)^{c}}\right] \;,
\end{eqnarray}
where ${\rm Li}_{c}\left({z}\right)=\sum_{l\geq 1}\frac{z^{l}}{l^{c}}$
is the polylogarithm function and $z^{*}$ is determined by the relation 
\begin{eqnarray}\label{eq_zstar}
{\rm Li}_{c}\left({z}^*\right)=e^{\beta\Delta} \;.
\end{eqnarray}
This result (\ref{eq_TIDSI:P1_TgtTc}) implies that in the paramagnetic phase, the average value of the largest domain scales logarithmically with $L$,  
$\left\langle l_{\max}\right\rangle \approx b_L \approx a\,\log L$. In this phase, the PDF of the $l_n$'s has an exponential tail and besides, 
the typical number of domains is $\propto L$. Therefore, the fact that the limiting distribution is given by a Gumbel law (\ref{eq_TIDSI:P1_TgtTc}), which is known to describe the EVS of i.i.d. variables with an exponential tail \cite{gumbel1958statistics} (see also Appendix \ref{app:evs}), shows that the correlations among the $l_n$'s, generated by the global constraint in Eq. (\ref{eq:jpdf}), do not play any role for $T>T_c$. Note that a similar property was found for renewal processes with exponentially distributed intervals in Ref. \cite{GMS2015}.

\item $T=T_{c}$: along the critical line, the marginal distribution of the domain size has an algebraic tail $P(l) \sim l^{-c}$. In this case, we find that depending on the value of $c$ ($c>2$ or $1<c<2$), the PDF of $l_{\max}$ exhibits two different behaviors. 

\begin{itemize}
\item[(i)] If $c>2$, we show that the limiting distribution is asymptotically given by a Fr\'echet
distribution (see also Fig. \ref{fig_critical_c3} below): 
\begin{equation}
P_{1}(x|L)\approx\exp\left(-\left(d\,\frac{x}{L^{\frac{1}{c-1}}}\right)^{1-c}\right) \;, \; d = \left[\frac{1}{(c-1)\zeta(c-1)}\right]^{1/(1-c)} \;. \label{eq_TIDSI:P1_TeqTc_cgt2}
\end{equation}
Therefore in this case, the average value of $l_{\max}$ grows algebraically (and sub-linearly) with $L$, $\left\langle l_{\max}\right\rangle \propto L^{\frac{1}{c-1}}$. Besides, for $c>2$, the number of domains is still extensive, $\approx L/\zeta(c-1)$ \cite{bar2014mixed,bar2014mixed2} and therefore the limiting distribution found here (\ref{eq_TIDSI:P1_TeqTc_cgt2}) coincides with the result of EVS for i.i.d. random variables with an algebraic PDF \cite{gumbel1958statistics} (see also Appendix \ref{app:evs}), demonstrating that in this case the global constraint on the $l_n$'s in Eq.~(\ref{eq:jpdf}) is irrelevant. This is also in line with the results found for renewal processes in~Ref.~\cite{GMS2015}.

\begin{table}[tt]
\centering
\begin{tabular}{|c||c|c|c|c|}
\hline
&&&&\\
$\qquad$ & $T>T_c$ & $T=T_c$ & $T=T_c$ &$T<T_c$\\
$\qquad$&$\qquad$&$c\geq 2$ (II \& III) &$1<c<2$ (I)&$\qquad$\\
\hline\hline
&&&&\\
$\langle l_{\max}\rangle \approx$&$a_{\rm para}\,\log L$& $a_{\rm crit}\,L^{1/(c-1)}$ & $A_1 \, L$ & $L$\\
&&&&\\
\hline
&&&&\\
Cumulative dist. $P_1(x|L) \approx$ &$F_1\left[(x-b_L)/{a}\right]$&$F_2\left[x/L^{1/(c-1)}\right]$&$F_3(x/L)$& $F_4(L-x)$ \\
&&&&\\
$\quad $& Gumbel & Fr\'echet & $\neq$ i.i.d. case & $\neq$ i.i.d. case \\
&&&&\\
\hline
\end{tabular}
\caption{Summary of the main results for the average value $\langle l_{\max}\rangle$ and its cumulative distribution $P_1(x|L)$ in the different regions of the phase diagram depicted in Fig. \ref{fig:Phase_diagram}. The amplitudes are given by $a_{\rm para} = a$ [see Eq. (\ref{expr:a})], $a_{\rm crit} = d^{-1}\,\Gamma[(c-2)/(c-1)]$ [see Eq. (\ref{eq_TIDSI:P1_TeqTc_cgt2})] and $A_1\equiv A_1(c)$ is given in Eq. (\ref{eq:av_lmax}). The functions $F_1(y), F_2(y), F_3(y)$ and $F_4(y)$ -- which are different in these four different cases -- can be read off from Eqs. (\ref{eq_TIDSI:P1_TgtTc}), (\ref{eq_TIDSI:P1_TeqTc_cgt2}), (\ref{eq_TIDSI:P1_TeqTc_clt2}-\ref{eq_TIDSI:P1_TeqTc_clt2_lap}) and (\ref{eq:scaling_ferro}-\ref{eq:ferro_GF}) respectively.}  
\label{tab1}
\end{table}

\item[(ii)] If $c<2$, the statistics of $l_{\max}$ is quite different from the predictions of EVS for i.i.d. random variables. First we show that, in this case, $l_{\max} \sim L$ and in particular its first moment is given by
\begin{eqnarray}\label{eq:av_lmax}
\langle l_{\max}\rangle \sim A_1(c) \, L \;, \; A_1(c) = \frac{1}{c-1} \int_0^\infty \frac{\Gamma(1-c,x)}{\Gamma(1-c,x)-\Gamma(1-c)} dx \;,
\end{eqnarray}
where $\Gamma(\alpha,z) = \int_z^\infty x^{\alpha-1} e^{-x} \, dx$ is the incomplete gamma function. Note that $-\Gamma(1-c) > 0$ for $1<c\leq 2$.
Besides we show that in this case the limiting PDF of $l_{\max}$ is given by (see Fig. \ref{Fig:critical_c13} below)
\begin{equation}
P_{1}\left(x|L\right)\approx1-H_1\left(\frac{L}{x}\right),\label{eq_TIDSI:P1_TeqTc_clt2}
\end{equation}
where $H_1(u)$, which is defined for $u\geq 1$, obeys the following relation
\begin{equation}\label{eq_TIDSI:P1_TeqTc_clt2_lap}
\int_0^\infty e^{-wu} H_1(u) \, u^{c-2}du= \frac{\Gamma(c-1)}{w^{c-1}} \frac{\Gamma(1-c,w)}{\Gamma(1-c,w)-\Gamma(1-c)} \;.
\end{equation}
The function $H_1(u)$ is a piece-wise analytic function, which has singularities at all integer values of $u > 1$ (while $H_1(u) = 0$ for $u\leq 1$). In particular, for $1<u<2$, $H_1(u)$ can be computed explicitly 
\begin{equation}
H_1\left(u\right) = B(c) \; u^{2-c}\left(u-1\right)^{2c-2}\ _{2}F_{1}\left(1,c,2c-1,1-u\right) \;, \; 1<u<2 \;, \label{eq_TIDSI:P1_TeqTc_clt2_1ltult2}
\end{equation}
with $_{2}F_{1}$ being the hypergeometric function and $B(c) = -\Gamma(c-1)/[\Gamma(1-c)\Gamma(2c-1)] > 0$. One can also check from Eq. (\ref{eq_TIDSI:P1_TeqTc_clt2_lap}) that $H_1(u) \to 1$ as $u \to \infty$, as it should (as $P_1(x|L) \to 0$ when $x\to 0$). For instance, for the special case $c=3/2$, $H_1(u) = \sqrt{u}-1$, for $1<u<2$. The asymptotic behaviors of the PDF $p_1(x|L)$, in the large $L$ limit, are given in Eqs. (\ref{eq:g}) and  (\ref{eq:asympt_gy}). The non-trivial distribution $H_1(u)$ in Eq.~(\ref{eq_TIDSI:P1_TeqTc_clt2_lap}) (see also Fig. \ref{Fig:critical_c13} below) indicates that the global constraint is important in this case. The extensivity of $\langle l_{\max} \rangle \propto L$ together with its non trivial PDF, exhibiting non analytic behaviors, is reminiscent of the results found for renewal processes, as in Eq. (\ref{eq:renewal}), when $f(l)$ exhibits heavy tails \cite{FIK95,GMS2009,GMS2015} -- corresponding here to $c\leq 2$.  

\hspace*{0.5cm}In this regime, one may wonder whether the largest domain is the only extensive
one, or whether other domains are extensive. To answer this question, we have computed
the statistics of the $k^{th}$ largest domain,
$l_{\max}^{(k)}$. We found that $l_{\max}^{(k)}$ is extensive for any finite $k$. In particular, its average is given by
\begin{eqnarray}\label{eq:exprlk}
\langle l^{(k)}_{\max} \rangle \sim A_k(c) \, L \;, \; A_k(c) = \frac{1}{c-1} \int_0^\infty \left[\frac{\Gamma(1-c,x)}{\Gamma(1-c,x)-\Gamma(1-c)}\right]^k dx \;,
\end{eqnarray} 
while its cumulative distribution $P_k(x|L) = \Pr.(l_{\max}^{(k)} \leq x)$ reads, for large $L$,
\begin{eqnarray}
P_{k}(x|L) & \approx & 1-H_{k}\left(\frac{L}{x}\right),\label{eq_TIDSI:Pk}\\
\int_0^\infty\frac{e^{-wu}}{u^{2-c}}H_{k}(u)du & = & \frac{\Gamma(c-1)}{w^{c-1}} \left(\frac{\Gamma(1-c,w)}{\Gamma(1-c,w)-\Gamma(1-c)} \right)^k \nonumber \;,
\end{eqnarray}
which, for $k=1$, yields back the formula in Eq. (\ref{eq_TIDSI:P1_TeqTc_clt2_lap}). These results imply that
for $c<2$, there are, at the critical point, many macroscopic domains. Note that from Eq. (\ref{eq:exprlk}) one easily checks that $\sum_{k=1}^\infty \langle l^{(k)}_{\max}\rangle = L$. Besides, from Eq. (\ref{eq_TIDSI:Pk}), one can show that $H_k(u) = 0$ for $0<u\leq k$ (as the $k$-th largest domain is necessarily {\it smaller} than $L/k$), while $H_k(u) \to 1$ as $u \to \infty$. As for $k=1$, one can also show that $H_k(u)$ has singularities at every integer values of $u\geq k$. Note that the $k$-th longest excursion for renewal processes (\ref{eq:renewal}), and $f(l) \sim l^{-3/2}$, was recently studied in Ref. \cite{SV2015}.

\end{itemize}
\item $T<T_{c}$: in this case it is more convenient to focus on the PDF $p_1(x|L) = P_1(x|L) - P_1(x-1|L) = \Pr.( l_{\max} = x)$. We find that, for large $L$, keeping $x$ fixed, it reads
\begin{eqnarray}\label{eq:scaling_ferro}
p_1(x|L) \approx f_{\rm ferro} (y= L-x) \;,
\end{eqnarray}
where the generating function of the scaling function $f_{\rm ferro}(y)$, with $y \in {\mathbb N}$, is given by (see also Fig. \ref{Fig_ferro} below)
\begin{eqnarray}\label{eq:ferro_GF}
\sum_{y=0}^\infty z^y f_{\rm ferro}(y) = \left( \frac{1-e^{-\beta \Delta} \zeta(c)}{1-e^{-\beta \Delta} {\rm Li}_c(z)}\right)^2 \;.
\end{eqnarray}
The asymptotic behaviors of $f_{\rm ferro}(y)$ can easily be extracted from this expression (\ref{eq:ferro_GF}) and they are given in Eq. (\ref{eq:asympt_ferro}) below. Note that Eq. (\ref{eq:scaling_ferro}) implies that for large $L$, the maximum domain size is given
by $l_{\max}\approx L$. Besides the limiting distribution $f_{\rm ferro}(x)$ is actually quite different from the standard
limiting distributions known from the EVS of i.i.d. random variables, which shows that the  
global constraint among the $l_n$'s (\ref{eq:jpdf}) is actually important in the ferromagnetic phase. Interestingly, the 
limiting distribution $f_{\rm ferro}(x)$ has an algebraic tail [see Eq. (\ref{eq:asympt_ferro})], $f_{\rm ferro}(x) \propto x^{-c}$. This 
indicates that $\langle l_{\max} \rangle - L \approx {\cal O}(1)$ for $c\geq 2$ while $\langle l_{\max} \rangle - L \sim {\cal O}(L^{2-c})$ for $1< c < 2$ (and a logarithmic growth for $c=2$). In this case, one can show that the size of the next maxima $l^{(k)}_{\max}$, for $k\geq 2$ are all of order $1$, $l^{(k)}_{\max} \approx {\cal O}(1)$. Their distribution, that depends on $k$, is rather cumbersome and is not given here.

\end{itemize}

To summarize, the general picture is that in the paramagnetic phase
domains are small, and hence correlations -- which emerge due to
the global constraint (\ref{eq_TIDSI:sum_rule}) -- are essentially negligible.
In the ferromagnetic phase the maximal domain consists of almost all
of the sites of the chain, and the typical fluctuations are of order ${\cal O}(1)$.
At the transition, if $c>2$ the domains are again small (sub-extensive)
and the effect of correlations is negligible, but for $c<2$ the maximal
domain is extensive and the correlations are relevant. These results are summarized in Table~\ref{tab1}.

\subsection{Derivation of the results}

The starting point of our analytical computations is an exact expression for the cumulative distribution $P_{1}\left(x | L\right) = \Pr.\left(l_{\max}\leq x\right)$ of the largest domain $l_{\max}=\max_{1\leq n \leq N} l_{n}$
in the TIDSI model. It is simply obtained by summing up the joint PDF of the domains in Eq. (\ref{eq:jpdf}) over the lengths $l_n$'s from $1$ to $x$ (for a fixed value of the number $N$ of domains) and then by summing over all possible values of $N$. This yields the following ratio:
\begin{equation}
P_{1}\left(x|L\right) = \frac{W_0(x|L)}{Z(L)},\label{eq_TIDSI:P1}
\end{equation}
where $Z(L)$ is the partition function given in Eq. (\ref{eq:partition_func}) and $W_0(x|L)$ is thus given by
\begin{eqnarray}
W_0(x | L) & = & \sum_{N=1}^\infty\sum_{l_{1}=1}^{x}...\sum_{l_{N}=1}^{x}\prod_{n=1}^{N}\frac{e^{-\beta\Delta}}{l_{n}^{c}}\delta_{\sum_{n=1}^{N}l_{n},L} \;.\label{eq_TIDSI:ZLx}
\end{eqnarray}
Obviously, $Z(L) = \lim_{x \to \infty} W_0(x,L)$. Similarly, to compute the cumulative distribution of the $k$-th largest domain $l_{\max}^{(k)}$, $P_{k}\left(x | L\right) = \Pr\left(l_{\max}^{(k)}\leq x\right)$, it is useful to first introduce an auxiliary probability $W_p(x|L)/Z(L)$, with an integer $p\geq 0$, which denotes the probability that there are exactly $p$ domains whose size are bigger than $x$. For the event that the $k$-th largest domain has length less than or equal to $x$ to occur, there must be {\it at most} $k-1$ domains with lengths bigger or equal to $x$ (see for instance Ref. \cite{SM2012}).   
The cumulative probability $P_k(x|L)$ can then be written as 
\begin{equation}\label{eq:starting_Pk}
P_{k}(x|L)=\frac{1}{Z(L)}\sum_{p=0}^{k-1}W_{p}\left(x|L\right) \;,
\end{equation}
where $W_0(x,L)$ is given in Eq.~(\ref{eq_TIDSI:ZLx}) while, for $p \geq 1$, $W_p(x|L)$ is computed straightforwardly as
\begin{equation}
W_{p}\left(x|L\right)=\sum_{N=p}^{\infty}e^{-N\beta\Delta}\binom{N}{p}\sum_{l_{1}=x+1}^{\infty}...\sum_{l_{p}=x+1}^{\infty}\sum_{l_{p+1}=1}^{x}...\sum_{l_{N}=1}^{x}\prod_{n=1}^{N}\frac{1}{l_{n}^{c}}\delta_{\sum_{n=1}^{N}l_{n},L} \;, \;\; p \geq 1 \label{eq_TIDSI:Wkx0x1}
\end{equation}
where the binomial coefficient $\binom{N}{p}$ is a simple combinatorial factor counting the number of different ways to choose these $p$ largest domains among $N$.  

We start by analyzing the distribution of the largest domain, $P_{1}(x|L)$,
above, at and below the critical temperature $T_c$. The difficulty with evaluating expressions such as (\ref{eq_TIDSI:ZLx})
and (\ref{eq_TIDSI:Wkx0x1}) comes from the constraint over the domain
sizes. To handle such sums, it is customary, see for instance \cite{bar2014mixed2,evans2008condensation}, to work with the corresponding generating functions with respect to (w.r.t.) $L$ (in the language of statistical physics, this amounts to shift from the canonical to the grand-canonical ensemble). One obtains
\begin{eqnarray}
\widetilde{W}_0(x,z) & = & \sum_{L=1}^{\infty}W_0(x|L)z^{L}\label{eq_TIDSI:Qxz} \nonumber \\
 & = & \sum_{N=1}^\infty\prod_{n=1}^{N}\left(\sum_{l=1}^{x}\frac{e^{-\beta\Delta}z^{l}}{l^{c}}\right)=\frac{e^{-\beta\Delta}{\Phi}_{c}\left(z,x\right)}{1-e^{-\beta\Delta}{\Phi}_{c}\left(z,x\right)} \;,\label{eq_TIDSI:Qxz_sol}  
\end{eqnarray}
where the function $\Phi_{c}(z,x)$ is given by
\begin{eqnarray}\label{eq:def_Phi}
\Phi_{c}(z,x) & = & \sum_{l=1}^{x}\frac{z^{l}}{l^{c}} \;.
\end{eqnarray}
These explicit and exact formulae in Eqs. (\ref{eq_TIDSI:Qxz_sol}) and (\ref{eq:def_Phi}) are our starting point to extract the large $L$ behavior
of $W_0(x|L)$, via the Cauchy's inversion formula
\begin{equation}
W_0\left(x|L\right)=\frac{1}{2\pi i}\oint\frac{1}{z^{L+1}}\widetilde{W}_0\left(x,z\right)dz,\label{eq_TIDSI:cauchy}
\end{equation}
where the integration contour runs around the origin and does not contain any singularities of
$\widetilde{W}_0\left(x,z\right)$. Eventually one obtains $P_1(x|L)$ from Eq. (\ref{eq_TIDSI:P1}), in the different regions of the phase diagram. 

Similarly, the generating function of $W_p(x|L)$ can also be expressed as
\begin{eqnarray}\label{eq:tildeWp_1_text}
\widetilde W_p(x,z)& = & \sum_{L=1}^{\infty}W_{p}\left(x|L\right)z^{L} \nonumber \\
 & = & \sum_{N=p}^{\infty}e^{-N \beta\Delta}\binom{N}{p}\left[{\rm Li}_{c}\left(z\right)-\Phi_{c}\left(z,x\right)\right]^{p}\left[\Phi_{c}\left(z,x\right)\right]^{N-p} \;,\; p \geq 1 \; \;,
\end{eqnarray}
where ${\rm Li}_c(z) = \lim_{x \to \infty} \Phi_c(z,x)$ denotes the polylogarithm function
\begin{eqnarray}\label{def:polylog}
{\rm Li}_c(z) = \sum_{l=1}^\infty \frac{z^l}{l^c} \;.
\end{eqnarray}
It is straightforward to perform the sum over $N$ in Eq. (\ref{eq:tildeWp_1_text}) to obtain 
\begin{eqnarray}
\label{eq:tildeWp_2}
\widetilde W_p(x,z)& = & \frac{e^{-p\beta\Delta}\left[{\rm Li}_c(z) - \Phi_c(z,x)\right]^p}{\left[1 - e^{-\beta \Delta} \Phi_c(z,x) \right]^{p+1}} \;, \;  p \geq 1 \;.
\end{eqnarray}
Note that this expression is valid only for $p\geq 1$, while for $p=0$ Eq. (\ref{eq_TIDSI:Qxz_sol}) holds. Finally, $W_p(x|L)$ for $p \geq 1$ can also be obtained via Cauchy's inversion formula
\begin{equation}
W_p\left(x|L\right)=\frac{1}{2\pi i}\oint\frac{1}{z^{L+1}}\widetilde{W}_p\left(x,z\right)dz,\label{eq_TIDSI:cauchy_p}
\end{equation}
where the integration contour runs around the origin and does not contain any singularities of
$\widetilde{W}_p\left(x,z\right)$. Eventually one obtains $P_k(x|L)$ from Eq. (\ref{eq:starting_Pk}), in the different regions of the phase diagram.

\subsubsection{The largest domain in the paramagnetic phase $\left(T>T_{c}\right)$}

In this regime $e^{-\beta \Delta} \zeta(c) < 1$ [see Eq. (\ref{eq_TIDSI:Tc})] and to compute $W_0(x,L)$ from Eq. (\ref{eq_TIDSI:cauchy}), for large $L$, one notices that, for fixed $x$, $\widetilde{W}_0\left(x,z\right)$ has a simple pole at $z^*(x)$ [see Eq. (\ref{eq_TIDSI:Qxz_sol})]
given by 
\begin{equation}
1-e^{-\beta\Delta}\Phi_{c}\left(z^{*}(x),x\right)=0 \;.\label{eq_TIDSI:zstar}
\end{equation}
For $T>T_{c}$, $z^{*}(x)<1$ for all finite $x$ [this can be checked from Eq. (\ref{eq_TIDSI:Tc})]. Because of the existence of this pole the
integral in (\ref{eq_TIDSI:cauchy}) can be evaluated to leading order
in the large $L$ limit as 
\[
W_0\left(x|L\right)\approx\left[z^{*}(x)\right]^{-L} \;.
\]
In particular, $Z(L) = \lim_{x \to \infty} W_0(x|L) = (z^*)^{-L}$ where $z^* \equiv z^*(x \to \infty)$. Using that $\Phi_c(z,x \to \infty) = {\rm Li}_c(z)$, Eq.~(\ref{eq_TIDSI:zstar}) implies that $z^*$ satisfies Eq. (\ref{eq_zstar}). And therefore
\begin{equation}
P_{1}\left(x\right) = \frac{W_0(x|L)}{Z(L)}\approx\left(\frac{z^{*}}{z^{*}\left(x\right)}\right)^{L} \;. \label{eq_TIDSI:P1_zstar}
\end{equation}
We now compute $z^*(x)$ for large $x$ from Eq. (\ref{eq_TIDSI:zstar}). This is done by using the large $x$ expansion of $\Phi_{c}\left(z,x\right)$:
\begin{eqnarray}\label{eq:phi_largex}
\Phi_{c}\left(z,x\right) = {\rm Li}_c(z) - \sum_{l=x+1}^\infty \frac{z^l}{l^c} &=& {\rm Li}_c(z) - z^{x+1} \sum_{l=0}^\infty \frac{z^l}{(x+1+l)^c} \nonumber \\
&=& {\rm Li_c}(z) - \frac{1}{1-z}\frac{z^{x+1}}{x^c}\left(1 + {\cal O}(x^{-1}) \right) \;.
\end{eqnarray} 
By using this asymptotic expansion (\ref{eq:phi_largex}), we find that $z^*(x)$, which is solution of Eq. (\ref{eq_TIDSI:zstar}), admits the large $x$ expansion:
\begin{eqnarray}\label{eq:zstar_largex}
z^*(x) = z^{*} + \frac{1}{1-z^*} \frac{1}{{\rm Li}_{c-1}(z^*)}  \frac{{z^*}^{x+2}}{x^c} \left( 1 + {\cal O}(x^{-1})\right) \;,
\end{eqnarray}
where we recall that $z^*$ satisfies Eq. (\ref{eq_zstar}). From Eq. (\ref{eq_TIDSI:P1_zstar}) together with Eq. (\ref{eq:zstar_largex}) one obtains:
\begin{eqnarray}\label{eq:P1_para}
P_1(x|L) \sim \left( 1 + \frac{1}{1-z^*} \frac{1}{{\rm Li}_{c-1}(z^*)}  \frac{{z^*}^{x+1}}{x^c}\right)^{-L} \;.
\end{eqnarray}
In the large $L$ limit, Eq. (\ref{eq:P1_para}) eventually leads to the Gumbel distribution announced in Eq. (\ref{eq_TIDSI:P1_TgtTc}). 

\begin{figure}[tt]
\centering
\includegraphics[angle=-90,width=0.55\linewidth]{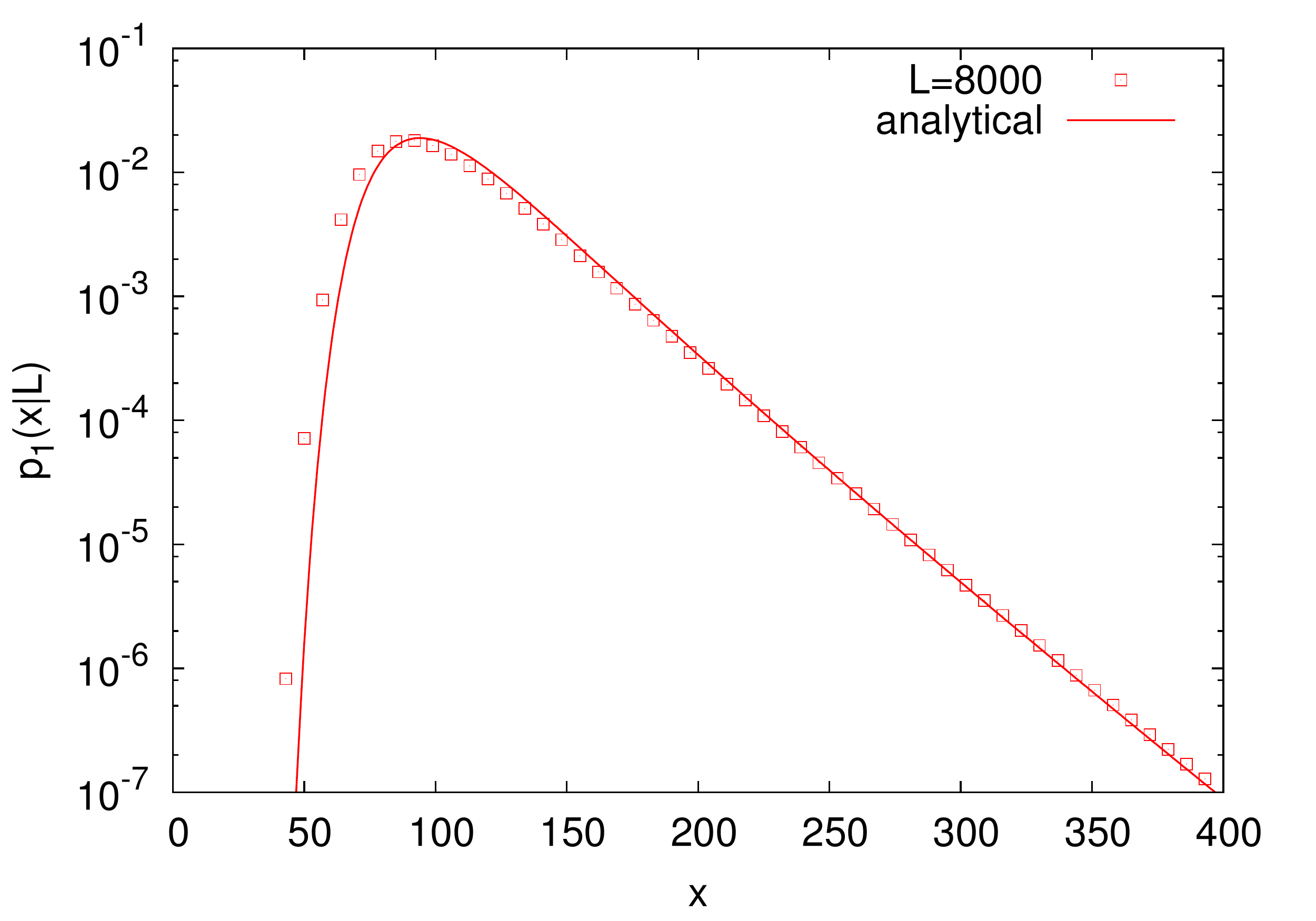}
\caption{Log-linear plot of $p_1(x|L)$ for $c=3/2$ and $e^{-\beta \Delta} = 0.5$ (hence in the paramagnetic phase). The square symbols correspond to a numerical evaluation of $p_1(x|L)$ for $L=8000$ (see Appendix \ref{app:numerics} for details) while the solid line corresponds to the exact formula given in Eq. (\ref{eq:P1_para}).}\label{Fig:plot_para}
\end{figure}
In Fig. \ref{Fig:plot_para} we show a comparison between a numerical estimate of $p_1(x|L)$ and the exact analytical formula derived from Eq. (\ref{eq:P1_para}). Note that the small discrepancy observed for small values of $x$ is a finite $L$ effect. For very large $L$, one expects that this formula (\ref{eq:P1_para}) converges to the Gumbel form given in Eq. (\ref{eq_TIDSI:P1_TgtTc}). Note that, for finite $L$, one expects to observe rather strong finite size effects, as this is the case for the EVS of i.i.d. random variables~\cite{GMOR2008}.

\subsubsection{The largest domain in the ferromagnetic phase $\left(T<T_{c}\right)$}

We start by evaluating the partition function $Z(L)$ in Eq. (\ref{eq:partition_func}). Its generating function $\tilde Z(z)$ is given by
\begin{eqnarray}
\widetilde{Z}(z) = \lim_{x \to \infty} \widetilde{W}_0(x,z) = \frac{e^{-\beta \Delta} {\rm Li}_c(z)}{1 - e^{-\beta \Delta} {\rm Li}_c(z)} \;,
\end{eqnarray}
from which one can show that the large $L$ behavior of $Z(L)$ is controlled by the branch point at $z=1$ of $\widetilde{Z}(z)$. Indeed, 
here, and in the following, we will use the asymptotic behavior of the polylogarithm function, 
\begin{eqnarray}
{\rm Li}_c(z) = \zeta(c) + (1-z)^{c-1}\left[\Gamma(1-c) + {\cal O}((1-z)^3) \right] + (1-z) \left[-\zeta(c-1) + {\cal O}(1-z) \right] \;. \label{eq:exp_polylog}
\end{eqnarray} 
From this asymptotic behavior (\ref{eq:exp_polylog}), one obtains the behavior of $\widetilde{Z}(z)$ for $z$ close $1$ as
\begin{eqnarray}\label{expansion_GF_Z_ferro}
\widetilde{Z}(z) \sim \frac{e^{-\beta \Delta}}{1-e^{-\beta \Delta}\zeta(c)} + \frac{e^{-\beta \Delta}}{(1-e^{-\beta \Delta}\zeta(c))^2} \Gamma(1-c) (1-z)^{c-1}(1 + {\cal O}(1-z)) + {\rm regular} \; {\rm terms} \;. 
\end{eqnarray} 
From Eq. (\ref{expansion_GF_Z_ferro}), one thus obtains the large $L$ behavior of $Z(L)$ as
\begin{eqnarray}\label{eq:asympt_Z}
Z(L) \approx \frac{e^{-\beta \Delta}}{(1-e^{-\beta \Delta}\zeta(c))^2} \, L^{-c} \;.
\end{eqnarray}
In the ferromagnetic phase, it is more convenient to compute the PDF of $l_{\max}$ (instead of the cumulative distribution), as in the case of ZRP \cite{evans2008condensation}. It reads
\begin{eqnarray}
p_1(x|L) &=& \Pr.(l_{\max} = x) = P_1(x|L) - P_1(x-1|L) = \frac{1}{Z(L)}\left[W_0(x|L)- W_0(x-1|L) \right]\;. 
\end{eqnarray}
Using the expression of $W_0(x|L)$ in Eqs. (\ref{eq_TIDSI:cauchy}) and (\ref{eq_TIDSI:Qxz_sol}) one obtains
\begin{eqnarray}\label{eq:pdf1}
p_1(x|L) = \frac{e^{-\beta \Delta}}{Z(L)} \frac{1}{2\pi i}\oint\frac{1}{z^{L+1}} \frac{z^x}{x^c} \frac{1}{1-e^{-\beta \Delta} \Phi_c(z,x)} \frac{1}{1-e^{-\beta \Delta} \Phi_c(z,x-1)} dz \;.
\end{eqnarray}  
Setting $x = L-y$ in Eq. (\ref{eq:pdf1}) one finds
\begin{eqnarray}\label{eq:pdf2}
p_1(L-y|L) &=& e^{-\beta \Delta} \frac{(L-y)^{-c}}{Z(L)} \frac{1}{2\pi i}\oint\frac{1}{z^{y+1}}  \frac{1}{1-e^{-\beta \Delta} \Phi_c(z,L-y)} \frac{1}{1-e^{-\beta \Delta} \Phi_c(z,L-y-1)} dz \;.
\end{eqnarray}
Therefore using the asymptotic behavior of $Z(L)$ in (\ref{eq:asympt_Z}), one obtains the limiting expression of $p_1(L-y|L)$ in Eq.~(\ref{eq:pdf2}), for fixed $y$ and $L \to \infty$ as
\begin{eqnarray}
p_1(L-y|L) \approx \left(1-e^{-\beta \Delta}\zeta(c)\right)^2 \frac{1}{2\pi i}\oint\frac{1}{z^{y+1}}  \frac{1}{\left[1-e^{-\beta \Delta} {\rm Li}_c(z)\right]^2} \, dz \;.
\end{eqnarray} 
Therefore one has
\begin{eqnarray}\label{eq:ferro_explicit}
p_1(x|L )=p_1(L-y|L) \approx f_{\rm ferro}(y) = \left(1-e^{-\beta \Delta}\zeta(c)\right)^2 \frac{1}{2\pi i}\oint\frac{1}{z^{y+1}}  \frac{1}{\left[1-e^{-\beta \Delta} {\rm Li}_c(z)\right]^2} \, dz \;. 
\end{eqnarray}
This is equivalent to the expression given in Eq. (\ref{eq:ferro_GF}), upon using the Cauchy's inversion formula. It is easy to derive now the asymptotic behavior of the function $f_{\rm ferro}(y)$ from Eq. (\ref{eq:ferro_GF}). For example, taking the limit $z \to 0$ in Eq. (\ref{eq:ferro_GF}) and using ${\rm Li}_c(0)=0$, one obtains $f_{\rm ferro}(0) = (1-e^{-\beta \Delta} \zeta(c))^2$. In contrast, by taking the $z\to 1$ limit and using the asymptotic properties of ${\rm Li}_c(z)$ in Eq. (\ref{eq:exp_polylog}), it is easy to show that $f_{\rm ferro}(y)\sim A_{\rm ferro}/y^c$ for large $y$ (where $L$ has been sent
to infinity already), where $A_{\rm ferro} = 2e^{-\beta \Delta}/(1-e^{-\beta \Delta}\zeta(c))$. The asymptotic behaviors of $f_{\rm ferro}(y)$ can thus be summarized as follows
\begin{eqnarray}\label{eq:asympt_ferro}
f_{\rm ferro}(y) \approx
\begin{cases}
&\left(1- e^{-\beta \Delta} \zeta(c) \right)^2 \;, \; y \to 0 \;, \\
& \\
& \dfrac{A_{\rm ferro}}{y^c} \;, \; y \to \infty\;.
\end{cases}
\end{eqnarray}

\begin{figure}[t]
\centering
\includegraphics[angle=-90,width = 0.55\linewidth]{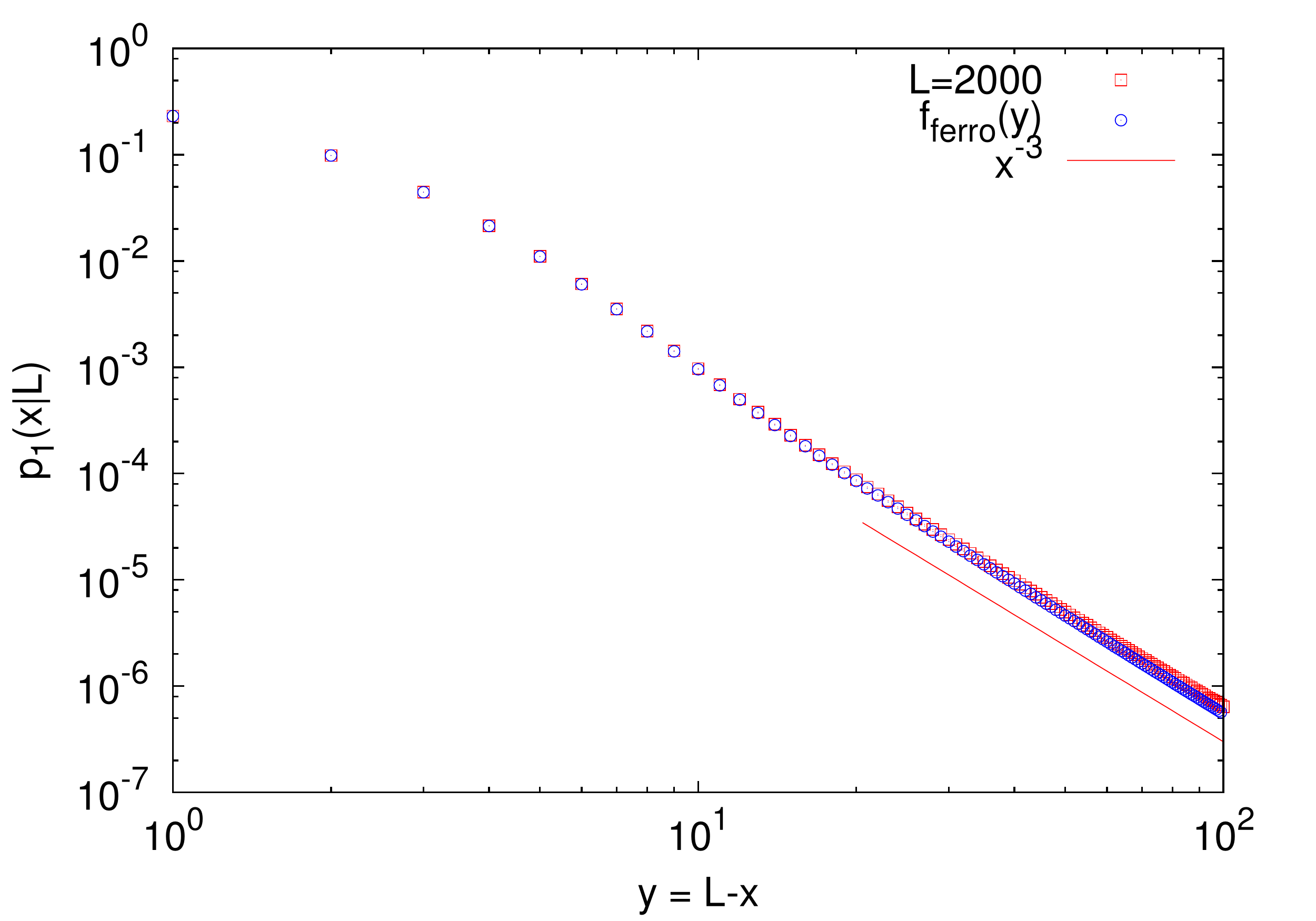}
\caption{Plot of $p_1(x|L)$ for $c=3$ and $e^{-\beta \Delta} = 0.2$ (hence in the ferromagnetic phase). The square symbols correspond to a numerical evaluation of $p_1(x|L)$ for $L=2000$ (see Appendix \ref{app:numerics} for details) while the circular symbols correspond to the exact limiting distribution $f_{\rm ferro}(y)$ obtained by expanding the right hand side of Eq. (\ref{eq:ferro_GF}) in powers of $z$. The slight discrepancy between the numerical and the exact asymptotic results is a finite $L$ effect. The solid line is a guide to the eyes indicating the expected algebraic behavior $\propto x^{-c}$, see Eq.~(\ref{eq:asympt_ferro}).}\label{Fig_ferro}
\end{figure}

In Fig. \ref{Fig_ferro} we show a comparison between a numerical evaluation of $p_1(x|L)$ for large $L=2000$ and the exact asymptotic result in Eqs. (\ref{eq:scaling_ferro}) and (\ref{eq:ferro_GF}).

\subsubsection{The largest domain at the critical point ($T=T_{c}$), for $1<c<2$}

We recall that in this case $e^{\beta_c \Delta} = \zeta(c)$, see Eq. (\ref{eq_TIDSI:Tc}). In this case the pole $z^{*}\left(x\right)<1$ for finite $x$ converges to the branch-cut at $z=1$ as $x\rightarrow\infty$, which then dominates the integral in Eq. (\ref{eq_TIDSI:cauchy}). We thus set $z=e^{-s}$, and evaluate $\Phi_{c}\left(e^{-s},x\right)$ when $s \to 0$. It is then convenient to rewrite $\Phi_{c}\left(e^{-s},x\right)$ as 
\begin{eqnarray}
\Phi_{c}\left(e^{-s},x\right) & = & \sum_{l=1}^{x}\frac{e^{-s l}}{l^{c}} \nonumber \\
& = & {\rm Li}_c(e^{-s})-\sum_{l=x+1}^{\infty}\frac{e^{-s l}}{l^{c}} \;. \label{eq:decomposition}
\end{eqnarray}
We recall the asymptotic behavior of the polylogarithm function [see Eq. (\ref{eq:exp_polylog})], 
\begin{eqnarray}
{\rm Li}_c(e^{-s}) = \zeta(c) + s^{c-1}\left[\Gamma(1-c) + {\cal O}(s^3) \right] + s \left[-\zeta(c-1) + {\cal O}(s) \right] \;, \label{eq:exp_polylog_bis}
\end{eqnarray}
whose leading behavior thus depends on whether $c<2$ or $c>2$. Besides, in the limit $s \to 0$, the discrete sums over $l$ in Eq. (\ref{eq:decomposition}) can be replaced, to leading order, by integrals. Therefore, in the limit $s \to 0$, $x \to \infty$ keeping $sx$ fixed one obtains (using Eq. (\ref{eq:exp_polylog_bis}) for $c<2$ here) 
\begin{eqnarray}\label{Phi_small_s}
\Phi_{c}\left(e^{-s},x\right) & \approx & \zeta(c) + s^{c-1}\left[\Gamma\left(1-c\right)-\Gamma\left(1-c,sx\right)\right] \;,
\end{eqnarray}
where we recall that $\Gamma(\alpha,z) = \int_z^\infty x^{\alpha-1} e^{-x} \, dx$. Using this small $s$ behavior (\ref{Phi_small_s}) in the expression for $\widetilde{W}_0(x,e^{-s})$ in Eq. (\ref{eq_TIDSI:Qxz_sol}), one obtains 
\begin{eqnarray}\label{eq:tildeW_0_final}
\widetilde W_0(x,e^{-s}) \sim \frac{\zeta(c)}{s^{c-1}} \frac{1}{\Gamma(1-c,sx)-\Gamma(1-c)} \;, \; {\rm for} \;\; s \to 0 \;, x \to \infty \;, \; {\rm keeping} \; sx  \;\; {\rm fixed} \;,
\end{eqnarray}
where we have used $e^{-\beta_c \Delta} \zeta(c) = 1$, see Eq.~(\ref{eq_TIDSI:Tc}). This formula in Eq. (\ref{eq:tildeW_0_final}), evaluated in the limit $x \to \infty$ (for finite $s$)  yields, using $Z(L) = \lim_{x \to \infty} W_0(x|L)$
\begin{eqnarray}\label{eq:laplace_Z}
\sum_{L=1}^\infty e^{-s L} Z(L) \approx \int_0^\infty e^{-s L} Z(L) dL \approx - \frac{\zeta(c)}{\Gamma(1-c)} s^{1-c} \;, 
\end{eqnarray}
which yields the large $L$ behavior of the partition function $Z(L)$
\begin{eqnarray}\label{eq:ZL}
Z(L) \approx - \frac{\zeta(c)}{\Gamma(1-c) \Gamma(c-1)} L^{c-2} \;, \; L \to \infty \;.
\end{eqnarray}
On the other hand, by rewriting the small $s$ behavior of $\widetilde W_0(x,e^{-s})$ in Eq. (\ref{eq:tildeW_0_final}) as
\begin{eqnarray}\label{eq:tildeW_0simpl}
\widetilde W_0(x,e^{-s}) = -\frac{\zeta(c)}{\Gamma(1-c)s^{c-1}}\left(1 - \frac{\Gamma(1-c,sx)}{\Gamma(1-c,sx)-\Gamma(1-c)} \right)
\end{eqnarray}
we obtain that $W_0(x|L)$ takes the following scaling form, for $L \to \infty$, $x \to \infty$ keeping $x/L$ finite:
\begin{eqnarray}\label{eq:W0_scaling}
W_0(x|L) \sim L^{c-2} \left[-\frac{\zeta(c)}{\Gamma(1-c)\Gamma(c-1)} \right]\left(1 - H_1\left(\frac{L}{x}\right) \right) \;.
\end{eqnarray} 
Dividing by $Z(L)$ given in Eq. (\ref{eq:ZL}) one then gets
\begin{eqnarray}\label{eq:P1_text}
P_1(x|L) = \frac{W_0(x|L)}{Z(L)} = 1 - H_1\left(\frac{L}{x} \right) \;,
\end{eqnarray}
where the scaling function $H_1(L/x)$ satisfies
\begin{eqnarray}\label{eq:Laplace:discrete}
\sum_{L=1}^\infty e^{-sL} L^{c-2} H_1\left( \frac{L}{x}\right) \approx \frac{\Gamma(1-c,sx)}{\Gamma(1-c,sx)-\Gamma(1-c)}  \;.
\end{eqnarray}
%
%
This equation holds in the scaling limit where $L \to \infty$, $x\to \infty$ keeping the ratio $L/x$ fixed. Equivalently, in the Laplace space,  
this corresponds to taking $s \to 0$, $x \to \infty$, keeping $sx$ finite. In the limit $s \to 0$, the discrete sum over $L$ can be replaced by an integral over the continuous variable $L$. Performing the change of variable $u = L/x$ in that integral yields the relation for $H_1(u)$ announced in Eq. (\ref{eq_TIDSI:P1_TeqTc_clt2_lap}):
\begin{equation}\label{eq_TIDSI:P1_TeqTc_clt2_lap_text1}
\int_0^\infty e^{-wu} H_1(u) \, u^{c-2}du= \frac{\Gamma(c-1)}{w^{c-1}} \frac{\Gamma(1-c,w)}{\Gamma(1-c,w)-\Gamma(1-c)} \;.
\end{equation}
It turns out that this expression can be inverted explicitly in the range $1< u < 2$ (see Appendix \ref{app:H1} for details). One obtains 
\begin{equation}
H_1\left(u\right) = B(c) \; u^{2-c}\left(u-1\right)^{2c-2}\ _{2}F_{1}\left(1,c,2c-1,1-u\right) \;, \; 1<u<2 \;, \label{eq_TIDSI:P1_TeqTc_clt2_1ltult2_text}
\end{equation}
with $_{2}F_{1}$ being the hypergeometric function and $B(c) = -\Gamma(c-1)/[\Gamma(1-c)\Gamma(2c-1)] > 0$, as announced in the introduction in Eq. (\ref{eq_TIDSI:P1_TeqTc_clt2_1ltult2}). In particular, for $u \to 1$, $H_1(u)$ behaves as
\begin{eqnarray}
H_1(u) \approx B(c) (u-1)^{2c-2} \;, \; u \to 1 \;. \label{eq:asympt_H1} 
\end{eqnarray}

From the cumulative distribution in Eq. (\ref{eq:P1_text}), one can obtain the PDF of $l_{\max}$ as 
\begin{eqnarray}\label{eq:p_1_deriv}
p_1(x|L) \approx \frac{\partial}{\partial x} \left[1-H_1\left(L/x \right) \right] = \frac{L}{x^2} H_1'\left(L/x \right) \;.
\end{eqnarray} 
Multiplying both sides by $L$ one gets 
\begin{eqnarray}\label{eq:g}
L \, p_1(x|L) \approx g\left(\frac{x}{L} \right) \;, \; {\rm where} \; \; g(y) = \frac{1}{y^2} H'_1\left( \frac{1}{y}\right) \;.
\end{eqnarray}
From the asymptotic behavior of $H_1(u)$ in Eq. (\ref{eq:asympt_H1}) when $u \to 1$, one obtains the behavior of $g(y)$ for $y \to 1$, as 
$g(y) \approx 2 B(c)(c-1) (1-y)^{2c-3}$. On the other hand, in the opposite limit $yÊ\to 0$, we need to investigate the large $u$ asymptotics of $H_1(u)$ in Eq. (\ref{eq_TIDSI:P1_TeqTc_clt2_lap_text1}). In this limit, we need to study the poles of $H_1(u)$, i.e., the zeros of $w^{c-1}[\Gamma(1-c,w)-\Gamma(1-c)]$, which are denoted by $s_k$. These zeroes are such that $s_{\pm k} = - \alpha_k + i \beta_k$ with a negative real part ($\alpha_k >0$, for all $k$) and $0<\alpha_0<\alpha_1<\alpha_2<\ldots$. Furthermore, $s_0 = - \alpha_0$ is the only real zero ({\it i.e.}, $\beta_0 = 0$) \cite{Buc1969}. Therefore in the large $u$ limit, one has $1-H_1(u) \propto e^{-\alpha_0 u}$  and the amplitude can be computed explicitly by evaluating
the residue of the integrand at $w = - \alpha_0$. Finally, the asymptotic behaviors of $g(y)$ can be summarized as follows 
\begin{eqnarray}\label{eq:asympt_gy}
g(y) \approx
\begin{cases}
&\gamma_0 \, e^{-\alpha_0/y} y^{c-4} \left(1 - \dfrac{2-c}{\alpha_0} y \right)Ê+ {\cal O}(e^{-\alpha_1/y})\;, \; y \to 0 \;, \\
& \\
&\gamma_1\, (1-y)^{2c-3} + {\cal O}((1-y)^{2c-2})\;, \; y \to 1 \;,
\end{cases}
\end{eqnarray}
where $\gamma_0 = \pi \alpha_0^2e^{-\alpha_0}/[(c-1)\sin(\pi (c-1))]$ and $\gamma_1= 2 (c-1)B(c)$ and where $-\alpha_0$ is the single negative real zero of $\Gamma(1-c,w)-\Gamma(1-c)$ as a function of real $w$. Note also that, by looking at the asymptotic behavior of $g(y)$ for $y \to 1$ in Eq. (\ref{eq:asympt_gy}), one observes that $c=3/2$ appears as a kind of "transition" point. For $c<3/2$, $g(y)$ is diverging when $y \to 1$ while it is vanishing for $c>3/2$. Exactly at $c=3/2$, $g(y) = 1/(2 y^{3/2})$ and in this case $g(y \to 1) = 1/2$.   
\begin{figure}[t]
\centering
\includegraphics[width = 0.35\linewidth,angle=-90]{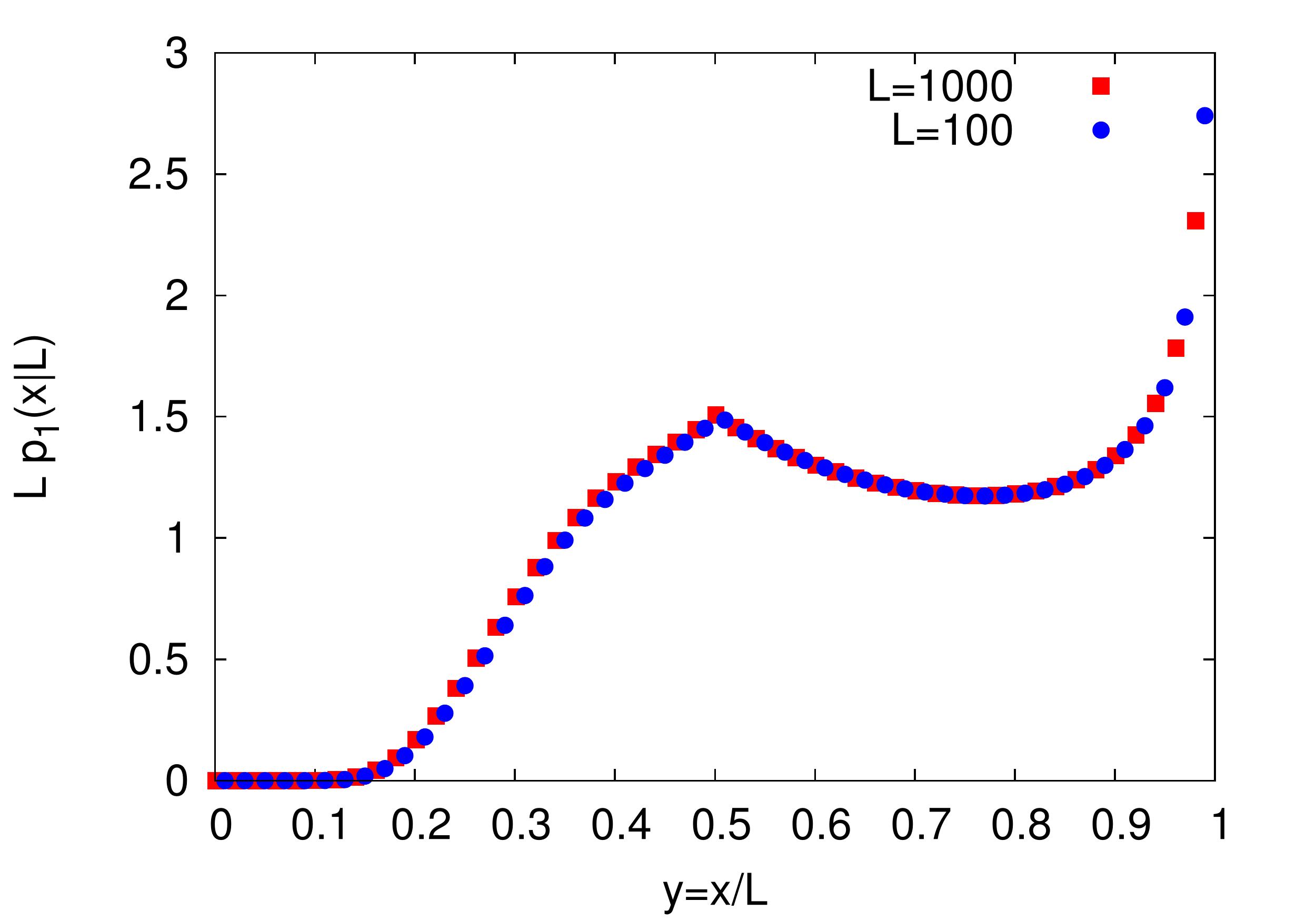}\includegraphics[width = 0.35\linewidth,angle=-90]{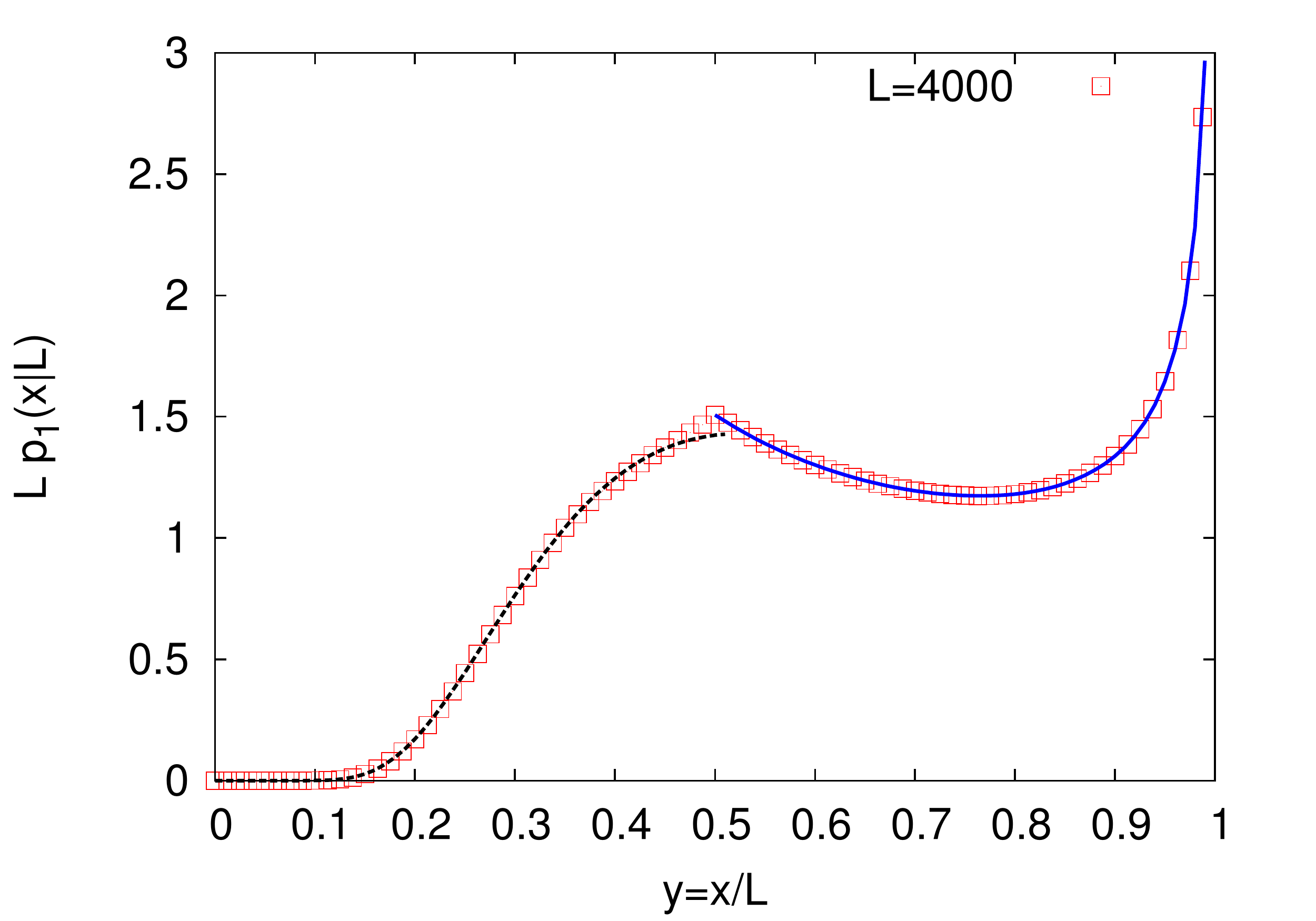}
\caption{{\bf Left:} Scaled plot of the PDF of $l_{\max}$, $L\,p_1(x|L)$ as a function of $x/L$ for $L=100$ and $L=1000$ at criticality and for $c=1.3$. The good collapse of the data (obtained from numerical simulations, [see Appendix \ref{app:numerics} for details]), for these two different values of $L$, corroborate the scaling form given in Eq. (\ref{eq:g}). The singularity for $x/L = 1/2$ is clearly visible on this plot, while other singularities (of higher order and hence not visible on this plot) also exist for $x/L = 1/k$, with $k=3,4, \ldots$. {\bf Right:} Scaled plot of the PDF of $l_{\max}$, $L\,p_1(x|L)$ as a function of $x/L$ for $L=4000$ at criticality and for $c=1.3$. The squared symbols correspond to a numerical evaluation of $p_1(x|L)$ while the blue solid line, for $x/L \geq 1/2$ corresponds to 
the exact result given in Eqs. (\ref{eq:g}) and (\ref{eq_TIDSI:P1_TeqTc_clt2_1ltult2_text}). The dotted line for $y< 1/2$ corresponds to the asymptotic behavior for $y \to 0$, given in Eq. (\ref{eq:asympt_gy}) (and in this case $\alpha_0 \approx 1.582$). }\label{Fig:critical_c13}
\end{figure} 

In Fig. \ref{Fig:critical_c13} (left panel) we show a scaled plot of the PDF of $l_{\max}$, $L\,p_1(x|L)$ as a function of $y=x/L$ for $L=100$ and $L=1000$ at criticality and for $c=1.3$. This plot shows a very good agreement with the scaling form predicted in  Eq. (\ref{eq:g}) $L p_1(x|L) \approx g(y=x/L)$. This plot also shows clearly the singularity of $p_1(x|L)$ for $x = L/2$, a feature which is commonly observed in the PDF of such extreme quantities in related models \cite{FIK95,GMS2009,GMS2015}. In the right panel of Fig. \ref{Fig:critical_c13} we show that our numerical data (for $L=4000$) are in very good agreement with our exact formula (the solid line) valid for $x/L \geq 1/2$, given in Eqs.~(\ref{eq:g}) and (\ref{eq_TIDSI:P1_TeqTc_clt2_1ltult2_text}). Furthermore, we show that the asymptotic behavior of $g(y)$ for $y \to 0$ in Eq. (\ref{eq:asympt_gy}) -- plotted as a dotted line in Eq. (\ref{eq:asympt_gy}) -- provides a very good estimate in the whole interval $[0,1/2]$.

To conclude this section, we compute the average value $\langle l_{\max}\rangle$, which is conveniently written as
\begin{eqnarray}\label{eq:lmax_1}
\langle l_{\max} \rangle = \sum_{x=1}^\infty [1 - P_1(x|L)] = \frac{1}{Z(L)} \sum_{x=1}^\infty [Z(L) - W_0(x|L)] \;.
\end{eqnarray}
From the results obtained in Eqs. (\ref{eq:laplace_Z}) and (\ref{eq:tildeW_0simpl}), one finds that the generating function of the numerator is given by
\begin{eqnarray}\label{eq:lmax_2}
\sum_{L=1}^\infty e^{-sL} \sum_{x=1}^\infty [Z(L) - W_0(x|L)] \approx -\frac{\zeta(c)}{\Gamma(1-c)} \frac{1}{s^{c}} \int_{0}^\infty \frac{\Gamma(1-c,u)}{\Gamma(1-c,u) - \Gamma(1-c)} \, du \;, \; s \to 0 \;.
\end{eqnarray}
These relations (\ref{eq:lmax_1}) and (\ref{eq:lmax_2}), together with the expression for $Z(L)$ in Eq. (\ref{eq:ZL}), lead to the result for $\langle l_{\max}\rangle$ announced in Eq. (\ref{eq:av_lmax}).

\subsubsection{The largest domain at the critical point ($T=T_{c}$), for $c>2$}

In this case, $\Phi_{c}\left(e^{-s},x\right)$ behaves, for $s \to 0$, as
\begin{eqnarray}
\Phi_{c}\left(e^{-s},x\right) & = & \sum_{l=1}^{x}\frac{e^{-s l}}{l^{c}} = {\rm Li}_c(e^{-s}) - \sum_{l=x+1}^\infty \frac{e^{-sl}}{l^c} \label{eq:polylog_cgt2} \;. 
\end{eqnarray}
For $c>2$, the first term behaves as ${\rm Li}_c(e^{-s}) = \zeta(c) - s \zeta(c-1) + o(s)$ [see Eq. (\ref{eq:exp_polylog})]. On the other hand, when $s \to 0$ the discrete sum over $l$ can be replaced by an integral over $l$ which, for $s\to 0$ and $x \to \infty$ behaves simply as
\begin{eqnarray}
\sum_{l=x+1}^\infty \frac{e^{-sl}}{l^c} \approx \frac{x^{1-c}}{c-1} \;.
\end{eqnarray}
Therefore one has the asymptotic behavior
\begin{equation}\label{eq:asympt:Phi:cgeq2}
\Phi_{c}\left(e^{-s},x\right)\approx\zeta(c)-s\zeta(c-1)-\frac{x^{1-c}}{c-1} \;.
\end{equation}
It thus follows from Eqs. (\ref{eq_TIDSI:Qxz_sol}) and (\ref{eq:asympt:Phi:cgeq2}) that 
\begin{eqnarray}
\widetilde W_0(x,e^{-s}) \approx \frac{\zeta(c)}{\zeta(c-1)} \frac{1}{s + d\,x^{1-c}}
\end{eqnarray}
where $d = 1/[(c-1)\zeta(c-1)]$, from which it follows that, for large $L$, 
\begin{eqnarray}\label{W0_cgeq2}
W_0(x|L) \sim e^{- d\, L x^{1-c}} \;.
\end{eqnarray}
The partition function $Z(L) = W_0(x\to \infty|L) \approx 1$ for large $L$ and $c>2$. Hence $P_1(x|L) = W_0(x|L)/Z(L)$ tends to the Fr\'echet 
distribution announced in Eq. (\ref{eq_TIDSI:P1_TeqTc_cgt2}).

\begin{figure}[ht]
\centering
\includegraphics[width = 0.4\linewidth,angle = -90]{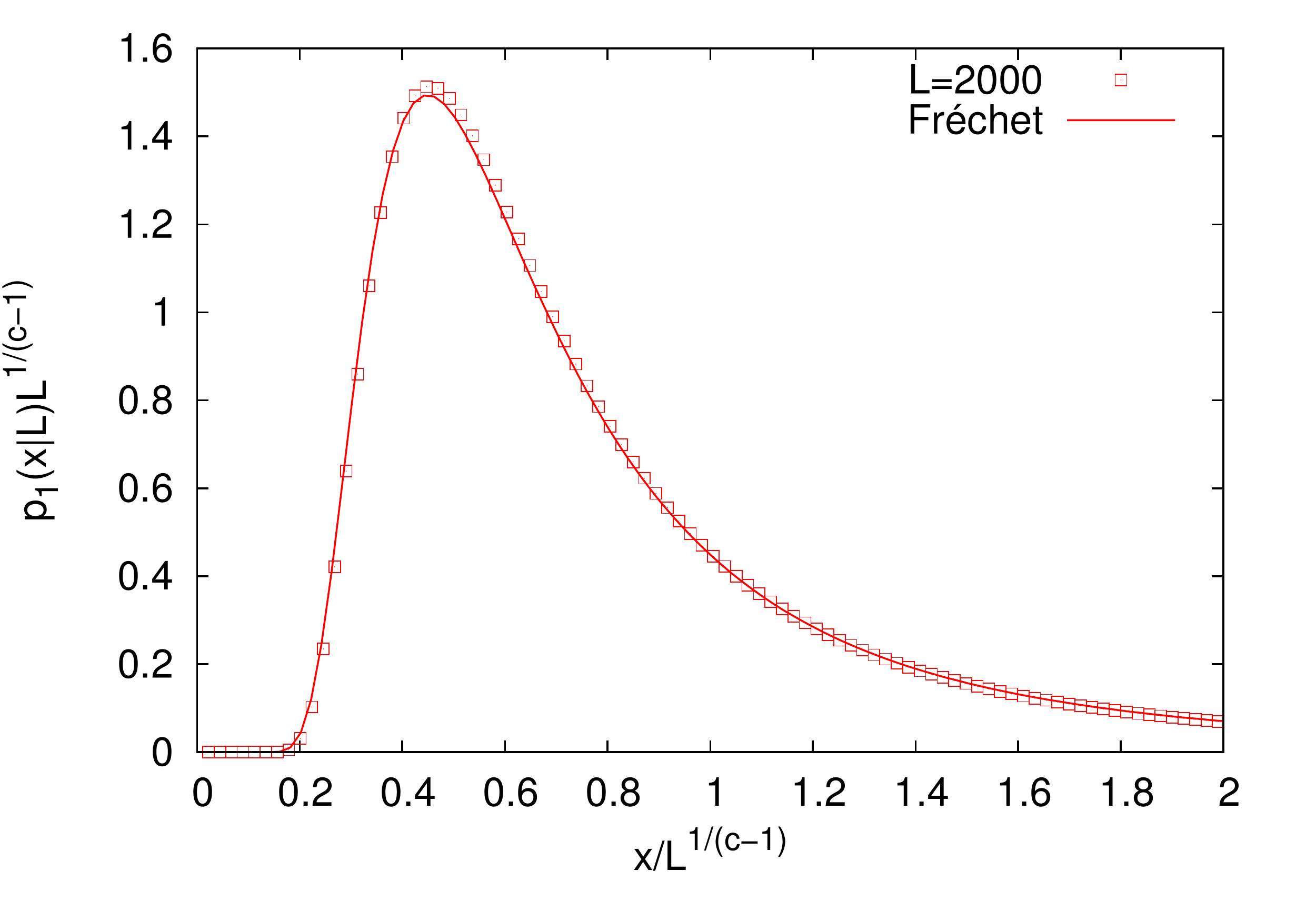}
\caption{Scaled plot of $p_1(x|L) \, L^{1/(c-1)}$ as a function of $x/L^{1/(c-1)}$, for $L=2000$, according to Eq. (\ref{eq_TIDSI:P1_TeqTc_cgt2}) -- we recall that $p_1(x|L) = P_1(x|L)-P_1(x-1|L)$ -- for $c=3$ at the critical temperature [see Eq. (\ref{eq_TIDSI:Tc})]. The solid line corresponds to the Fr\'echet distribution with parameters specified in Eq. (\ref{eq_TIDSI:P1_TeqTc_cgt2}).}\label{fig_critical_c3}
\end{figure}

In Fig. \ref{fig_critical_c3}, we show a scaled plot of the PDF of $l_{\max}$, $p_1(x|L)\, L^{1/(c-1)}$ as a function of $x/L^{1/(c-1)}$ evaluated numerically for $L=2000$. We see that the data are quite well described by our exact analytical prediction given in Eq. (\ref{eq_TIDSI:P1_TeqTc_cgt2}).   

\subsubsection{The $k^{th}$ largest domain}

The starting point for the analysis of the cumulative distribution $P_k(x|L)$ of the $k$-th largest domain is the expression given given in Eq. (\ref{eq:starting_Pk}) in terms of the probability $W_p(x|L)/Z(L)$ (\ref{eq_TIDSI:Wkx0x1}) that there are exactly $p$ domains whose size are larger than $x$. As done before for $P_1(x|L)$, it is convenient to study the generating function of $W_p(x|L)$ with respect to $L$, which reads, for $p \geq 1$
\begin{eqnarray}\label{eq:tildeWp_1}
\widetilde W_p(x,z)& = & \sum_{L=1}^{\infty}W_{p}\left(x|L\right)z^{L} \nonumber \\
 & = & \sum_{N=p}^{\infty}e^{-N \beta\Delta}\binom{N}{p}\left[{\rm Li}_{c}\left(z\right)-\Phi_{c}\left(z,x\right)\right]^{p}\left[\Phi_{c}\left(z,x\right)\right]^{N-p} \;.
\end{eqnarray}
It is straightforward to perform the sum over $N$ to obtain 
\begin{eqnarray}
\label{eq:tildeWp_2}
\widetilde W_p(x,z)& = & \frac{e^{-p\beta\Delta}\left[{\rm Li}_c(z) - \Phi_c(z,x)\right]^p}{\left[1 - e^{-\beta \Delta} \Phi_c(z,x) \right]^{p+1}} \;.
\end{eqnarray}

This formula (\ref{eq:tildeWp_2}) is exact in the whole phase diagram and we now focus on the critical line, where $T=T_c$, and restrict our attention to the case $c<2$ where one expects that $P_k(x)$ will be described by a non-trivial distribution, i.e. different from the one predicted by the EVS of i.i.d. sequences. In this case, we set $z= e^{-s}$ and the large $L$ behavior of $W_p(x|L)$ is governed by the behavior of $\widetilde W_p(x,e^{-s})$ when $s \to 0$. In the scaling limit, $s \to 0$, $x \to \infty$, keeping the product $sx$ fixed, one obtains, using the expansion of $\Phi_c(e^{-s},x)$ in Eq. (\ref{Phi_small_s}) together with ${\rm Li}_c(e^{-s}) \approx \zeta(c)+ \Gamma(1-c)s^{c-1}$:
\begin{eqnarray}
\label{eq:tildeWp_3}
\widetilde W_p(x,e^{-s})& \approx & \frac{\zeta(c)}{s^{c-1}} \frac{[\Gamma(1-c,sx)]^p}{[\Gamma(1-c,sx)-\Gamma(1-c)]^{p+1}} \;.
\end{eqnarray}
Note that setting $p=0$ in this formula (\ref{eq:tildeWp_3}) yields back the result obtained above for $\widetilde W_0(x,e^{-s})$ in the same scaling limit (\ref{eq:tildeW_0_final}). The probability $P_k(x)$ is obtained by summing up the probabilities $W_p(x|L)/Z(L)$ for $0\leq p \leq k-1$. From Eq. (\ref{eq:tildeWp_3}), one obtains straightforwardly
\begin{eqnarray}\label{eq:tildeWp_4}
\sum_{p=0}^{k-1} \widetilde W_p(x,e^{-s}) = -\frac{\zeta(c)}{s^{c-1}\Gamma(1-c)} \left(1- \left[\frac{\Gamma(1-c,sx)}{\Gamma(1-c,sx)-\Gamma(1-c)}\right]^k  \right) \;.
\end{eqnarray}
From this expression (\ref{eq:tildeWp_4}) together with Eq. (\ref{eq:starting_Pk}), by performing the same manipulations as for $k=1$, see Eqs. (\ref{eq:tildeW_0simpl})-(\ref{eq:Laplace:discrete}), one arrives at the expression for $P_k(x|L)$ given in Eq. (\ref{eq_TIDSI:Pk}). Finally, the computation of the average value $\langle l_{\max}^{(k)}\rangle$ can be performed along the same lines as for $k=1$ [see Eqs. (\ref{eq:lmax_1}), (\ref{eq:lmax_2})]. Using Eq. (\ref{eq:tildeWp_4}), this yields the result announced in Eq. (\ref{eq:exprlk}).

\section{Relation to other models}\label{section:other}

\subsection{Relation to the inverse distance squared Ising model}

As mentioned in the introduction, MOTs appear in several apparently
unrelated contexts. One of the first models which was studied in this
context is the one-dimensional Ising model with interactions decaying
as $r^{-2}$, named here the inverse distance squared Ising (IDSI)
model. The IDSI is defined, much like the TIDSI, on a spin chain of
size $L$, with the Hamiltonian
\begin{eqnarray}
\mathcal{H} & = & -\sum_{i<j}J\left(i-j\right)\sigma_{i}\sigma_{j},\label{eq_TIDSI:IDSI_Hamiltonian1}\\
J(r) & \approx & Cr^{-2}.\label{eq_TIDSI:IDSI_Hamiltonian2}
\end{eqnarray}
Thouless \cite{thouless1969long} was the first to suggest that this
model exhibits a discontinuous phase transition at some finite temperature,
{\it i.e.} that the magnetization changes discontinuously from $0$ to $m_{c}$
at the transition. Later Yuval and Anderson \cite{yuval1970exact,anderson1970exact}
used a scaling analysis to predict that the transition is critical.
Their analysis predicted that the correlation length has an essential
singularity at the transition,
\begin{eqnarray}\label{eq:xi_IDSI}
\xi\sim\exp\left[{\cal O}\left(\frac{1}{\sqrt{T-T_{c}}}\right)\right] \;.
\end{eqnarray}
Only a decade and a half later the mixed order nature of the transition
was proved rigorously by Aizenman {\it et al.}~\cite{aizenman1988discontinuity}. 

In addition to the structural similarity between Eqs. (\ref{eq_TIDSI:SC_Hamiltonian1})-(\ref{eq_TIDSI:SC_Hamiltonian2})
and Eqs. (\ref{eq_TIDSI:IDSI_Hamiltonian1})-(\ref{eq_TIDSI:IDSI_Hamiltonian2}),
the IDSI model exhibits mixed order symmetry breaking transition point,
which is qualitatively similar to the transition in the TIDSI model. However, the quantitative features of
this transition, namely the jump of the magnetization from $0$ to
$m_{c}<1$, and the essential singularity in the correlation length,
as well as more subtle features, are different from the TIDSI, as
discussed in \cite{bar2014mixed,bar2014mixed2}.

In terms of EVS, the authors are not aware of any systematic study
of the extreme value theory of the IDSI model. However, it is known
\cite{aizenman1988discontinuity} that domains, as defined in the
TIDSI, are microscopic in the IDSI model for any positive temperature.
At the transition there is a macroscopic structure that emerges, but
it is hidden and can be revealed in a Random Cluster Model perspective.
Hence, it is plausible to guess that the extreme value statistics
may follow a Gumbel distribution. Anyway the EVS, whatever form it
has, will probably not show the features we discussed above for the
TIDSI model.

\subsection{Relation to the Poland-Scheraga Model}

The Poland-Scheraga (PS) model \cite{PS1966} is a prototypical model
for studying thermal denaturation of DNA molecules, which is the process
in which the two strands of the DNA molecule separate upon heating.
The PS model idealizes the DNA chain of size $L$ as a set of alternating
bound and denatured segments. The degrees of freedom are the lengths
of these segments $\left\{ l_{1}, l_2, \cdots,l_N\right\} $. Bound segments contribute
linearly to the energy, so that 
\begin{eqnarray}\label{eq:H_PS}
\mathcal{H}=E_{b} \sum_{n=1}^{N/2}\,l_{2n-1},
\end{eqnarray}
with the constraint $\sum_{n=1}^N l_{n}=L$. Here $E_{b}<0$ is the binding
energy, $N$ is the total number of segments and we assumed that the
first segment is bound. Denatured segments, also known as \emph{loops},
carry no energy, but instead contribute to the entropy of a configuration,
due to the flexibility of single stranded DNA. Treating the strands
in a loop as random walkers which must meet implies that the entropy
of a loop of size $l$ takes the form 
\begin{eqnarray}\label{eq:S_PS}
S=\tilde{\Delta}+sl-c\log l \;.
\end{eqnarray}
Here $\tilde{\Delta}$ and $s$ are constants which depend on the
geometry of the embedding space and the chemical properties of the DNA, and $c$
--- the loop exponent --- is a universal parameter which depends only
on dimensionality of the embedding space and topological properties
of the DNA such as self avoidance. Therefore the Boltzmann
weight of a specific configuration $\left\{ l_{1}, l_2, \cdots, l_N\right\} $ can be
written as
\begin{eqnarray}
e^{-\beta E_{b}l_{1}}\frac{e^{\tilde{\Delta}+sl_{2}}}{l_{2}^{c}}e^{-\beta E_{b}l_{3}}\frac{e^{\tilde{\Delta}+sl_{4}}}{l_{4}^{c}}...
\end{eqnarray}
This Boltzmann weight is equivalent to the one derived from an effective
Hamiltonian
\begin{equation}
\mathcal{H}_{eff}=\sum_{n=1}^{N/2}E_{b}l_{2n-1}-\frac{1}{\beta}\sum_{n=1}^{N/2}\left(sl_{2n}-c\log l_{2n}\right).\label{eq_TIDSI:PS_Heff}
\end{equation}
Comparing Eq. (\ref{eq_TIDSI:PS_Heff}) and Eq. (\ref{eq_TIDSI:Hamiltonian})
implies that the TIDSI can be presented as a variant of the PS model
in which all segments are loops (and then the linear term can be gauged
out). An important difference between the models is that in the PS
model $c$ is a universal parameter, while in the TIDSI $C$ is a
parameter in the Hamiltonian. Actually, as is discussed below, the
role of $c$ is played in the TIDSI by $\beta_{c}C$, where $\beta_{c}$
is the inverse critical temperature. Other than this difference, the
models are very similar, both in their definition and in their phenomenology.
From the phenomenological perspective, the main difference is that
in the TIDSI we consider the magnetization order parameter, for which
the Hamiltonian is symmetric. The natural order parameter of the PS
model is the fraction of loop base-pairs, which have no such symmetry.
Because of this difference, in the PS model, the regime $1<c<2$ is
considered to be a continuous transition, as the natural order parameters
are continuous, while in the TIDSI the corresponding regime ($1<\beta_{c}C<2$)
exhibits a mixed order transition. The reason is that the magnetization
in the paramagnetic phase is protected by symmetry and hence cannot
be different than $0$, while no such symmetry protects the order
parameters of the PS model (and the density of domains in the TIDSI
model). Another difference between the phase diagrams, is that in
the PS model, for $c<1$ there is no transition (and there is also
a condition on $s$ and $\tilde{\Delta}$). The TIDSI model, on the
other hand, supports a transition for any value of its parameters,
but the effective parameter $c=\beta_{c}C$ satisfies $c>1$.

From the perspective of extreme value theory, the EVS of loops in
the PS model should be very similar to the EVS of the TIDSI model.
Bound segments, however, are microscopic for any $T<T_{c}$, and hence
their EVS follow the Gumbel distribution.

\section{Conclusion}

To summarize, we have presented, in this paper, a thorough study of 
fluctuations of the size of the largest domain $l_{\max}$ in the TIDSI model. 
We found that above the critical temperature $T>T_c$ for
any $c >1$, and for $c>2$ also at criticality $T=T_c$, the asymptotic EVS is similar to  that
of i.i.d. variables, indicating that the correlations are effectively weak. However, for $c<2$ at $T_c$ as well as in the ferromagnetic phase $T<T_c$ for any $c>1$, we have found novel extreme value distributions, which we have computed exactly (see Table \ref{tab1} for a summary
of the main results).

Studying the extreme value statistics of correlated variables is an
active field of research, and a specifically intriguing avenue is
the EVS of variables at criticality. In this work we focused on a novel
aspect of this topic, namely the EVS at mixed order transitions. As discussed
in section IV, it will be interesting to test the universality of the results found here, by
studying EVS for different models exhibiting MOT phase transition. In particular, 
as mentioned above, the TIDSI model has many similarities
to the Poland-Scheraga (PS) model for DNA denaturation, and it is easy to extend the
results above for the EVS for the case of loop sizes in the PS model.
It would be therefore interesting to study experimentally whether
the actual loop sizes distribution resembles the EVS that was found
in this paper. A related question is how real world details, such
as base-pair heterogeneity \cite{coluzzi2007numerical} and topological
constraints \cite{bar2012denaturation} affect the EVS of loops.


\acknowledgments

We would like to thank Michael Aizenman, Or Cohen and Ori Hirchberg
for stimulating discussions. We thank the Galileo Galilei Institute (Florence)
for Theoretical Physics and the International Center for Theoretical Siences (Bangalore), where part
of this work was achieved, for hospitality. S. N. M. wants to thank the hospitality of the Weizmann Institute, during
a visit as the Weston visiting professor in 2014, where this work started. The support of the Israel Science Foundation (ISF) is gratefully acknowledged.

\appendix

\section{A brief reminder on EVS for i.i.d. random variables}\label{app:evs}

Given $N$ i.i.d random variables $\left\{ x_{i}\right\} _{i=1}^{N}$, the distribution of their maximum, $m=\max_{i}\left\{ x_{i}\right\} $, properly
shifted and scaled, converges in the large $N$ limit to one of three max-stable distributions, depending
on the tail of the parent PDF of $x_i$'s, \textbf{$p(x)$}: 
\begin{enumerate}
\item If the support of $p(x)$ has an upper cutoff $b$, such that $p(x) \sim (b-x)^{\alpha-1}$ when $x \to b$, with $\alpha > 0$, then the limiting cumulative distribution of the maximum is given by the Weibull distribution:
\begin{equation}
P_{1}\left(z\right)=\Pr\left(m\leq z\right)\approx\begin{cases}
\exp\left(-\left(\frac{b-z}{a_{N}}\right)^{\alpha}\right) & z<b\\
1 & z\ge b
\end{cases} \; ,\; N \to \infty \;,\label{eq_P3:Weibull}
\end{equation}
where $a_{N}$ depends on $p\left(x\right)$.
\item If the support of $p\left(x\right)$ is unbounded and if it has a power-law tail, $p\left(x\gg1\right)\sim x^{-\alpha-1}$,
then the limiting cumulative distribution of $m$ is given by the Fr\'echet distribution:
\begin{equation}
P_{1}\left(z\right)\approx\begin{cases}
0 & z\le 0\\
\exp\left[-\left(\frac{z}{a_{N}}\right)^{-\alpha}\right] & z>0
\end{cases} \;, \; N \to \infty \;, \label{eq_P3:Frechet}
\end{equation}
where $a_{N}$ depends on $p\left(x\right)$. 
\item If the support of $p\left(x\right)$ is unbounded and decays faster than any power-law for
large $x$, then the limiting cumulative distribution of $m$ is given by the Gumbel
distribution
\begin{equation}
P_{1}\left(z\right)\approx\exp\left(-\exp\left(-\left(\frac{z-b_{N}}{a_{N}}\right)\right)\right) \;, \; N \to \infty \;. \label{eq_P3:Gumble}
\end{equation}
Here again $a_{N}$ and $b_{N}$ depend on $p\left(x\right)$. 
\end{enumerate}

\section{Analysis of the function $H_1(u)$}\label{app:H1}

In this appendix, we study the function $H_1(u)$ and derive its explicit expression for $1<u<2$ given in Eq. (\ref{eq_TIDSI:P1_TeqTc_clt2_1ltult2}). We recall that $H_1(u)$ is defined by the following relation [see Eq. (\ref{eq_TIDSI:P1_TeqTc_clt2_lap})]:
\begin{equation}\label{eq_TIDSI:P1_TeqTc_clt2_lap_text}
\int_0^\infty e^{-wu} H_1(u) \, u^{c-2}du= \frac{\Gamma(c-1)}{w^{c-1}} \frac{\Gamma(1-c,w)}{\Gamma(1-c,w)-\Gamma(1-c)} \;,
\end{equation}
where $\Gamma(\alpha,z) = \int_z^\infty x^{\alpha-1} e^{-x} \, dx$ is the incomplete gamma function. To analyze $H_1(u)$ it is convenient to expand the right hand side of Eq. (\ref{eq_TIDSI:P1_TeqTc_clt2_lap_text}) and write the inverse Laplace transform as 
\begin{eqnarray}\label{eq:Bromwich}
H_1(u)\, u^{c-2} &=& \int_{a-i\infty}^{a+i\infty} \frac{dw}{2\pi i}\, e^{w u}
\frac{\Gamma(c-1)}{w^{c-1}} \frac{\Gamma(1-c,w)}{\Gamma(1-c,w)-\Gamma(1-c)}  \nonumber \\
&=& -\sum_{n=1}^\infty \int_{a-i\infty}^{a+i\infty} \frac{dw}{2\pi i} e^{wu}\frac{\Gamma(c-1)}{w^{c-1}}  \left[\frac{\Gamma(1-c,w)}{\Gamma(1-c)} \right]^n \;,
\end{eqnarray}
where $a >0$ such that the singularities of the integrand, namely the negative real axis which is a branch cut and $w=0$ which is a branch point (for non-integer values $c$) or a pole (for integer values of $c$), are to the left of the Bromwich contour. Apart from that, the integrand has no other singularity. On the other hand, for large complex $w$ it is easy to see that
\begin{eqnarray}\label{eq:asympt_gamma}
\Gamma(1-c,w) \sim \frac{e^{-w}}{w^c} \;, \; |w| \to \infty \;.
\end{eqnarray}
Therefore, this implies that
\begin{eqnarray}\label{eq:asympt_2}
\int_{a-i\infty}^{a+i\infty} \frac{dw}{2\pi i} e^{wu}\frac{\Gamma(c-1)}{w^{c-1}}  \left[\frac{\Gamma(1-c,w)}{\Gamma(1-c)} \right]^n = 0 \;, {\rm for} \; u<n \;,
\end{eqnarray}
since the integral over $w$ can be computed by closing the Bromwich contour to the right. Therefore, from Eqs. (\ref{eq:Bromwich}) and (\ref{eq:asympt_2}) one obtains (i) that $H_1(u) = 0$ for $u<1$ -- which is expected since $l_{\max} \leq L$ and (ii) that, for $1<u<2$, $H_1(u)$ is determined only by the term $n=1$ in Eq. (\ref{eq:Bromwich}). This yields, for $1<u<2$:

\begin{eqnarray}
H_1(u)\, u^{c-2} & = & - \frac{\Gamma(c-1)}{\Gamma(1-c)} \int_{a-i\infty}^{a+i\infty}\frac{e^{wu}}{w^{c-1}}\int_{w}^{\infty}\frac{e^{-\eta}}{\eta^{c}}d\eta dw \nonumber \\
 & = & - \frac{\Gamma(c-1)}{\Gamma(1-c)}\int_{a-i\infty}^{a+i\infty}\frac{e^{wu}}{w^{c-1}}\int_{0}^{\infty}\frac{e^{-\left(wz+w\right)}}{\left(wz+w\right)^{c}}wdzdw \nonumber\\
 & = & - \frac{\Gamma(c-1)}{\Gamma(1-c)} \int_{0}^{\infty}\int_{a-i\infty}^{a+i\infty}\frac{e^{w\left(u-z-1\right)}}{w^{2c-2}\left(z+1\right)^{c}}dzdw \nonumber\\
 & = & - \frac{\Gamma(c-1)}{\Gamma(1-c)}\int_{0}^{\infty}\frac{\left(u-z-1\right)^{2c-3}}{\Gamma\left(2c-2\right)\left(z+1\right)^{c}}\Theta\left(u-z-1\right)dz \nonumber\\
 & = & - \frac{\Gamma(c-1)}{\Gamma(1-c)\Gamma\left(2c-1\right)} \left(u-1\right)^{2c-2}\ _{2}F_{1}\left(1,c,2c-1,1-u\right)\;,
\end{eqnarray}
which yields the result announced in Eq. (\ref{eq_TIDSI:P1_TeqTc_clt2_1ltult2}). In particular, it behaves as $H_1(u) \sim u^{2c-1}$ when $u \to 1$ from above. 

The expansion in Eq. (\ref{eq:Bromwich}) shows that $H_1(u)$ has actually singularities at any integer values of $u$. One can indeed show, from Eq. (\ref{eq:Bromwich}) that $H_1(k+\epsilon) - H_1(k-\epsilon) \sim \epsilon^{(k+1)c-2}$, with $k$ an integer, as $\epsilon \to 0$. 

\section{Exact numerical evaluation of EVS for TIDSI}\label{app:numerics}

The simple structure of the TIDSI model allows us to calculate efficiently the distribution of its largest domain. The basic
object is the truncated partition function $W_0\left(x|L\right)$
which is the sum of weights of TIDSI configurations of size $L$ for
which the maximal domain is smaller than $x$. The formula for $W_0\left(x|L\right)$
is given in Eq.~(\ref{eq_TIDSI:ZLx}), and it can be used to derive
a recursion relation for $W_0\left(x|L\right)$. For any $x\ge1$ one has indeed
\begin{eqnarray}
W_0\left(x|L=0\right) & = & 1\label{eq_appB:Z0x}\\
W_0\left(x|L=1\right) & = & e^{-\beta\Delta}\label{eq_appB:Z1x}\\
W_0\left(x|L\right) & = & \sum_{l=1}^{\min(x,L)}W_0\left(x|L-l\right)e^{-\beta\Delta}l^{-c} \;. \label{eq_appB:ZLx}
\end{eqnarray}
Using this together with Eq. (\ref{eq_TIDSI:P1}), recalling that $Z(L) = \lim_{x \to \infty} W_0(x|L) = W_0(L|L)$ the distribution
of the largest domain of the TIDSI model $P_1(x|L)$ can be numerically evaluated in a straightforward way.

\bibliographystyle{unsrt}

\end{document}